\documentclass[amssymb,amsmath,superscriptaddress,footinbib]{revtex4}
\usepackage{amsmath,graphicx,subfigure,subeqnarray,fancyhdr,amsmath,multirow,color, mathrsfs,relsize}
\usepackage{natbib}

\begin{document}

\newcommand{\change}{\textcolor{black}}

\title{Synchronized flutter of two slender flags}
\author{J\'er\^ome Mougel}
\email{jerome.mougel@imft.fr}
\affiliation{LadHyX -- D\'epartement de M\'ecanique, CNRS -- Ecole Polytechnique, 91128 Palaiseau, France}
\affiliation{Institut de M\'{e}canique des Fluides de Toulouse, All\'{e}e Camille Soula,  31400 Toulouse, France}
\author{Olivier Doar\'e}
\email{olivier.doare@ensta-paristech.fr}
\affiliation{Unit\'{e} de M\'{e}canique, ENSTA Paristech, Chemin de la Huni\`{e}re, 91761 Palaiseau, France}
\author{S\'ebastien Michelin}
\email{sebastien.michelin@ladhyx.polytechnique.fr}
\affiliation{LadHyX -- D\'epartement de M\'ecanique, CNRS -- Ecole Polytechnique, 91128 Palaiseau, France}
\date{\today}
\begin{abstract}
The interactions and synchronization of two parallel and slender flags in a uniform axial flow are studied in the present paper by generalizing Lighthill's Elongated Body Theory (EBT) and Lighthill's Large Amplitude Elongated Body Theory (LAEBT) to account for the hydrodynamic coupling between flags. The proposed method consists in two successive steps, namely the reconstruction of the flow created by a flapping flag within the LAEBT framework and the computation of the fluid force generated by this nonuniform flow on the second flag. In the limit of slender flags in close proximity, we show that the effect of the wakes have little influence on the long time coupled-dynamics and can be neglected in the modeling. This provides a simplified framework extending LAEBT to the coupled dynamics of two flags. 
Using this simplified model, both linear and large amplitude results are reported to explore the selection of the flapping regime as well as the dynamical properties of two side-by-side slender flags. Hydrodynamic coupling of the two flags is observed to destabilize the flags for most parameters, and to induce a long-term synchronization of the flags, either in-phase or out-of-phase. 
\end{abstract}
\maketitle

\section{Introduction}
A flexible plate or filament may flap spontaneously in a uniform axial flow as a result of the competition between its internal rigidity, its inertia and the destabilizing fluid forces resulting from the deflection of the fluid particles by the deforming structure. This flutter or ``flag'' instability and  resulting flapping motion have received much interest as exemplified in the recent review of Ref.~\cite{shelley2011}. Beyond its academic interest or traditional applications, flag flutter has also recently been studied to extract energy from an incoming flow, for example by converting the flapping motion into an electric current using flags covered by electro-active materials~\citep{giacomello2011, doare2011}. 

Understanding the hydrodynamic coupling of multiple flags is critical in this context, in particular to assess how it affects the flapping properties, synchronization and more generally the collective performance of an assembly of piezoelectric flags. Two-dimensional soap-film experiments by Ref.~\cite{zhang2000} on two flexible filaments revealed an in-phase synchronization for small separation distances, and an out-of-phase synchronization at larger distances, a trend later supported by further experimental, theoretical and numerical studies~\citep{zhu2003,jia2007}.
More insight on the two-dimensional multiple flag dynamics was also gained \change{from experiments}~\citep{schouveiler2009}, linear stability analysis~\citep{michelin2009} and numerical simulations~\citep{alben2009, farnell2004,tian2011}. 

These studies all focus on the two-dimensional problem which is representative of the three-dimensional case only when the width of the flags is much larger than its length. 
\change{The coupling of multiple flags for arbitrary span, and in particular for slender flags, remains poorly documented despite its practical importance, for energy harvesting purposes for instance.  Direct numerical simulations of the fluid-solid systems are possible~\citep*{banerjee2015}, but their complexity and computational cost prohibit at the moment systematic parametric studies or optimization, and emphasize the need for reduced-order modeling of these interactions. 
The main objective of the paper is therefore to provide a simplified model in the slender body limit allowing both to give insights to the physical synchronization process and to provide a useful benchmark for subsequent works on the topic.
 }
 
 In potential flow, fluid forces on a single flapping slender flag can be computed as a reactive local force: an added momentum $m_a \boldsymbol{u}_n$ is associated to each slice of fluid normal to the structure's centerline, with $\boldsymbol{u}_n$ the local normal relative velocity of the solid with respect to the background flow and $m_a$ the added mass coefficient of the structure's cross-section; the reactive force results from changes in the momentum of the fluid advected along the deformed structure. This idea is at the heart of \emph{Elongated Body Theory} (EBT)~\citep{lighthill1960} and its generalization to nonlinear flapping dynamics, the \emph{Large Amplitude Elongated Body Theory} (LAEBT)~\citep{lighthill1971}. The powerful advantage of this method is its simplicity: the fluid force is expressed solely in terms of the local kinematics of the solid body. Extensions to this theory have recently been  proposed for three-dimensional body motions~\citep*{candelier2011} and weakly non-uniform background flows~\citep*{candelier2013}. 
 
The present article extends this approach to model the flapping dynamics of several slender flags and is organized as follows. Section~\ref{sec:model} presents the problem's geometry, the relevant parameters and structural model, and \S~\ref{sec:LAEBT} describes the method for computing the fluid forces in the presence of hydrodynamic interactions, based on a generalization of Lighthill's LAEBT to the case of two flags. The linear stability and mode selection of the two-flag configuration is analyzed in \S~\ref{sec:linear} and \S~\ref{sec:non_linear} focuses on flag synchronization in the saturated large-amplitude dynamics. Finally, \S~\ref{sec:conclusions} proposes a discussion of the problem and new opportunities for future work. 

\begin{figure}
\centering
\begin{tabular}{cc}
\includegraphics[width=5.5cm]{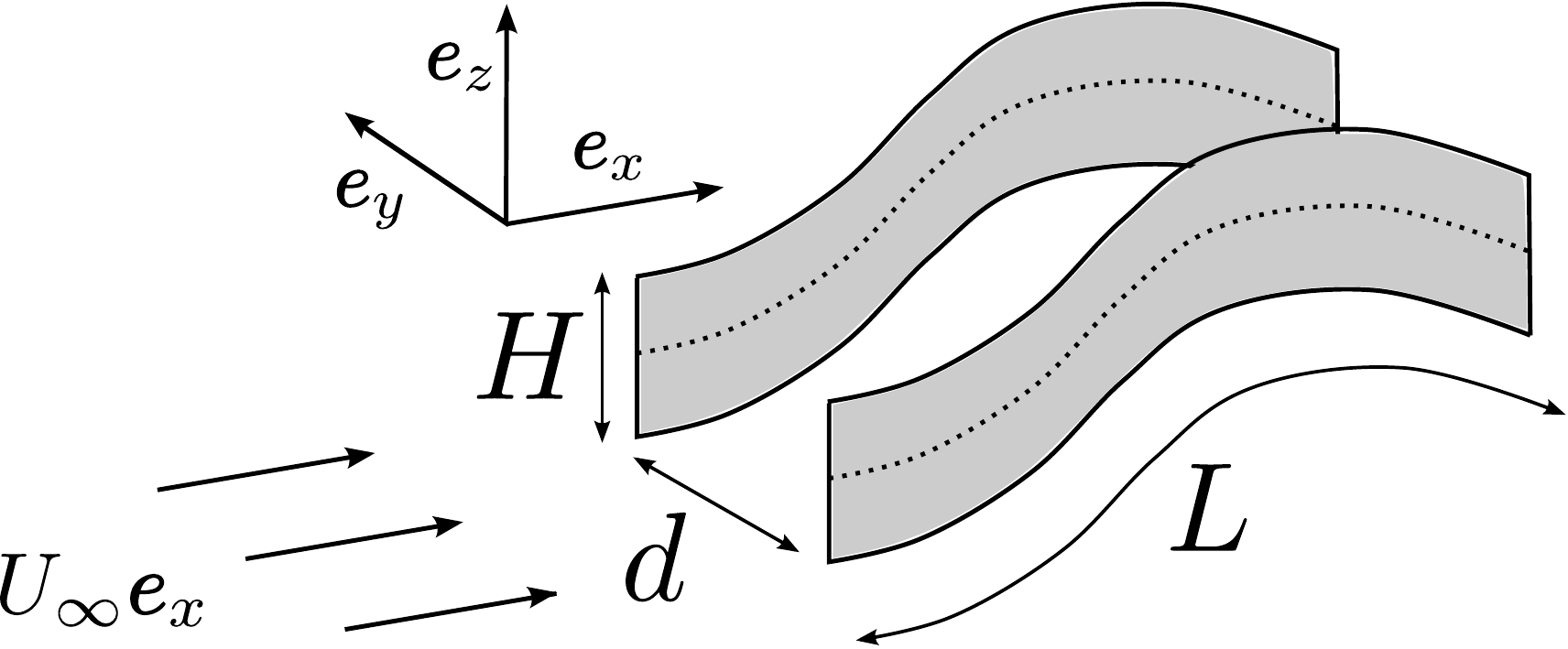} &
\includegraphics[width=6.5cm]{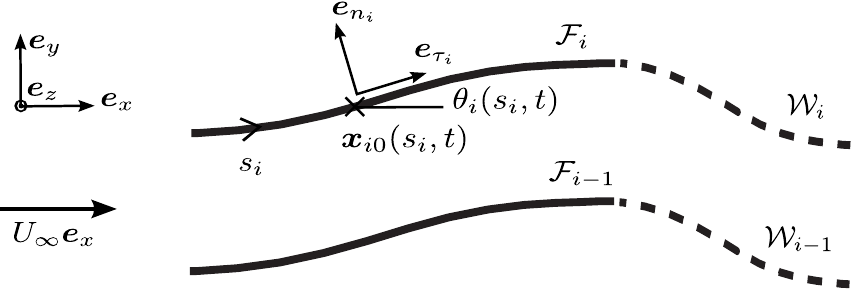} \\
$(a)$ & \hspace{1cm}  $(b)$
\end{tabular}
\caption{\label{fig:shema_flags} Sketches of the configuration: $(a)$ three-dimensional view, $(b)$ view from above.}
\end{figure}

\section{Problem setting}\label{sec:model}
We consider two parallel, rectangular and infinitely thin flags of length $L$ and width $H$ in a steady uniform flow of velocity $U_\infty$, density $\rho$ and kinematic viscosity $\nu$, with $d$ the distance between their clamped leading edges (Figure \ref{fig:shema_flags}$a$). Both flags are inextensible, with surface density $\mu$ and bending rigidity $B$; gravity is neglected. When $\mbox{Re}=U_\infty L/\nu\gg 1$, viscous effects are negligible except within thin boundary layers around the flags that separate at the trailing edge into vortex sheets; thus, a potential flow model is used. The flags' and flow dynamics are governed by four non-dimensional parameters, namely 
\begin{equation}
H^*= \frac{H}{L}, \hspace{0.5cm} d^*= \frac{d}{L}, \hspace{0.5cm} M^*= \frac{\rho L}{\mu}, \hspace{0.5cm} U^*=U_{\infty}L \sqrt{\frac{\mu H}{B}},
\end{equation}
which respectively correspond to the flags' aspect ratio, the non-dimensional inter-plate distance, the mass ratio and the reduced velocity. 

In the following, ${\cal F}_i$ and  ${\cal W}_i$ denote the $i^{th}$ flag and its wake ($i=1$, $2$).
Neglecting stream-wise torsion and span-wise displacement, the position $\boldsymbol{x}_i(s_i, z_i, t)$ of ${\cal F}_i$ is entirely described by its centerline position $\boldsymbol{x}_{i0}(s_i,t) = \boldsymbol{x}_{i}(s_i,z_i=0,t)$ with $s_i$ the Lagrangian curvilinear coordinate and $z_i$ the span-wise position. The local orientation of the centerline with respect to the incoming flow is $\theta_i(s_i,t)$, and $(\boldsymbol{e}_{\tau_i},\boldsymbol{e}_{n_i})$ denote the local tangent and normal unit vectors to the flag's surface (Figure \ref{fig:shema_flags}$b$). An Euler--Bernoulli beam model is used for each flag. Using $L$, $L/U_{\infty}$, $\rho H L^2$ as characteristic length, time and mass, the non-dimensional equations of motion for ${\cal F}_i$  read
\begin{align}
\label{eq:momentum}
 \frac{\partial^2 \boldsymbol{x}_{i0}}{\partial t^2} =  \frac{\partial}{\partial s_i} \left[ {f_T}_i {\boldsymbol{e}_{\tau_i}} - \frac{1}{U^{*2}}\frac{\partial^2 \theta_i}{\partial {s_i}^2} {\boldsymbol{e}_{n_i}}\right] + {\boldsymbol{f}_\textrm{fluid}}_i, 
\end{align}
\change{where ${f_T}_i(s_i, t)$ is the local tension, acting as a Lagrange multiplier to enforce each flag's inextensibility, namely}
\begin{align}
 \label{eq:inextensibility}
 \qquad{\boldsymbol{e}_{\tau_i}} =  \frac{\partial \boldsymbol{x}_{i0} }{\partial s_i},
\end{align}
and ${\boldsymbol{f}_\textrm{fluid}}_i(s_i, t)$ \change{is} the local \change{fluid} force applied on ${\cal F}_i$. 
The flags' coupling is purely hydrodynamic, and is therefore included in ${\boldsymbol{f}_\textrm{fluid}}_i$ which depends on the flags' kinematics, and is discussed in detail in the following.

\section{Fluid modelling}\label{sec:LAEBT}
\subsection{Preliminary discussion}
\label{subsec:order_magnitude}
In potential flow, the local pressure force on the flag is directly related to the local flow velocity which can be reconstructed using Biot--Savart law from the distribution of bound and free vorticity associated with ${\cal F}_i$ and ${\cal W}_i$.
 The relative magnitude of different hydrodynamic contributions can therefore be assessed by considering that of the induced flow velocity. 
In the case of two slender flags, we look for the dominant hydrodynamic terms depending upon $d$. 
This first approach \change{provides} a rough classification between the contributions of the neighbouring flag and that of the wakes. \change{Further quantitative justifications will be given in subsequent sections.} 
From Biot--Savart law, the contribution of the velocity field created on ${\cal F}_i$ by ${\cal F}_j$ is typically ${\cal O}(d^{-2})$ while, \change{away from the direct neighbourhood of the trailing edge}, the wakes contribute as ${\cal O}(L^{-2})$ for ${\cal W}_i$ (own wake) and  ${\cal O}((d^2+L^2)^{-1})$ for ${\cal W}_j$ (wake of the neighbour). Three cases must therefore be considered:
\begin{itemize}\renewcommand{\labelitemi}{$-$}
\item If $d\gg L$, the effect of ${\cal W}_i$ on ${\cal F}_i$ \change{is dominant over} the hydrodynamic coupling \change{between the flags}.
\item If $d = {\cal O}(L)$, all the contributions are of the same order and should all be retained.
\item If $d\ll L$, the effects of the wakes are negligible compared to hydrodynamic coupling. \change{More precisely a flag region of order ${\cal O} (H)$ near the trailing edge may feel a significant effect of the wakes. It will however be evidenced in the following that the global effect on the flapping dynamics nevertheless remains negligible in this limit. Note that this limitation is intrinsic to the original LAEBT and it is therefore consistent to try to extend this approach to the two-flags case in the range $d\ll L$.}
\end{itemize}

These first considerations indicate that it may be relevant to neglect wakes in the limit of flags in close proximity. In the following, we focus on the range $H\ll d \ll L$ for which the spirit of Lighthill's LAEBT can be extended naturally to account for hydrodynamic coupling: in that range, the effect of the wakes including their complex dynamics appears to remain negligible in front of the hydrodynamic coupling contribution. 

\subsection{Methodology}
For a single flag, an asymptotic expansion of the potential flow problem in the limit of small aspect ratio $H^*$ (but large displacement) provides the LAEBT formulation of a reactive force that depends exclusively on the local relative velocity of the flag to the background uniform flow~\citep{candelier2011}. This is particularly convenient as a detailed knowledge of the flow around the flag is unnecessary. Note also that this formalism reveals that the non-local effect of the wake is \change{negligible} in the slender body limit, and comes at higher order in the expansion in powers of $H^*$ as shown by \cite{eloy2010} in the linear case (EBT). For a freely-flapping body, the reactive force obtained here through LAEBT must be complemented by a local resistive force to account for lateral flow separation~\citep{eloy2012b,singh2012b}.

This local formulation is however lost for two flags, and the hydrodynamic perturbations induced by the second flag must be computed to determine ${\boldsymbol{f}_\textrm{fluid}}_i$. More specifically, in the limit of $H^*\ll d^*$, these flow perturbations remain subdominant in front of each flag's \change{dominant} self-contribution (at least while flags amplitudes remain small) and vary slowly along the flag's width.
The approach followed here is therefore to consider the motion of each flag within the weakly non-uniform local flow field created at its surface by its neighbour's motion. Two steps must be combined, namely (i) the reconstruction of the flow field created by a flapping flag, and (ii) a generalization of LAEBT (and of the resistive force) to account for non-uniformities and unsteadiness in the resulting local flow. These two points are detailed \change{below} and further combined to propose an extension of the LAEBT approach in the case of two slender flags.

\subsection{Flow created by a flapping flag in the LAEBT}
\label{subsec:flow_LAEBT}

In this section, an explicit expression of the flow created by a single flapping flag in the large-amplitude regime is obtained. \change{In potential flows, the velocity potential $\phi_i$ satisfies Laplace's equation in the fluid domain, that is $\Delta \phi_i=0$. In this framework, Green's second identity (see Ref.~\cite{jackson1999} for instance) leads to an expression of the velocity potential in the entire fluid domain from the knowledge of the velocity potential and its normal derivative (i.e. the normal flow velocity) on the flag and its wake.  As a consequence, the flow created in $\boldsymbol{x}$ by the $i^{th}$ flag and its wake reads
\begin{equation}
\label{eq:green_identity}
\phi_i(\boldsymbol{x},t) = \oint_{{\cal F}_i+{\cal W}_i}{ \left[ G(|\boldsymbol{x}-\boldsymbol{x}_i| ) \frac{\partial \phi_i(\boldsymbol{x}_i)}{\partial n_i}  -\phi_i(\boldsymbol{x}_i) \frac{\partial G(|\boldsymbol{x}-\boldsymbol{x}_i| ) }{\partial n_i}\,\right]dS_i(\boldsymbol{x}_i) },
\end{equation}
with  $G(r)=-1/(4 \pi r)$ the free-space Green function of Laplace equation and where integration should be performed here on both sides of the flag and wake.}

\change{In addition, the structure acts as an impermeable surface, so that the normal flow velocity matches that of the flag.
For infinitely thin structures, this leads to the continuity of the normal derivative of the flow potential, $\partial \phi(\boldsymbol{x}_i^+) /\partial n_i -\partial \phi(\boldsymbol{x}_i^-) /\partial n_i =0$, and the single-layer potential term in Eq.~\eqref{eq:green_identity} vanishes 
\begin{equation}
\label{eq:representation_th}
\phi_i(\boldsymbol{x},t) = -\int_{{\cal F}_i+{\cal W}_i}{[\phi_i](\boldsymbol{x}_i) \frac{\partial G(|\boldsymbol{x}-\boldsymbol{x}_i| ) }{\partial n_i}\,dS_i(\boldsymbol{x}_i)},
\end{equation}
where $[\phi_i]=\phi_i(\boldsymbol{x}_i^+) - \phi_i(\boldsymbol{x}_i^- )$ corresponds to the velocity potential jump across the flag and wake.}
In this formalism, the wake is assumed to consist of an infinitely thin vortical sheet of height $H$ extending to infinity.
Physically, Eq.~\eqref{eq:representation_th} corresponds to the flow induced by bound and free vorticity present in the vicinity of the flag and its wake respectively. 

In the slender body limit ($H<<L$), the potential jump is given by Ref.~\cite{candelier2011} and reads
\begin{equation}
\label{eq:pot_jump}
[\phi_i](\boldsymbol{x}_i)  = -2\tilde{u}_{n_i} \sqrt{{H^*}^2/4-{z_i}^2}.
\end{equation}
with $\tilde{u}_{n_i} = [\partial \boldsymbol{x}_{i0}/\partial t - \boldsymbol{e}_x]\cdot\boldsymbol{e}_{n_i}$.
Note that this elliptic form of the potential jump is analogous to the small-displacement limit (EBT, \citep{lighthill1970}); effectively, the large-amplitude case can locally be seen as a straight plate having normal relative velocity $\tilde{u}_{n_i}$. 

\change{
Using Eq.~\eqref{eq:pot_jump}, an explicit form of the flow created in the midplane, $z=0$, can be obtained from Eq.~\eqref{eq:representation_th} by integrating the contributions of the spanwise direction  (for $z_i$ from $-H^*/2$ to $H^*/2$):
\begin{equation}
\label{eq:potential_nff}
\phi_i(\boldsymbol{x},t) = -\int_{0}^{1}{ \frac{\tilde{u}_{n_i}(\boldsymbol{x}-\boldsymbol{x}_i) \cdot\boldsymbol{e}_{n_i}}{\pi [(x-x_i)^2 + (y-y_i)^2]^{1/2}} [E(X)-K(X)]\,ds_i(\boldsymbol{x}_i)} + \phi_{{\cal W}_i}(\boldsymbol{x},t,L_w).
\end{equation}
where $X=-{H^*}^2/(4[(x-x_i)^2 + (y-y_i)^2])$ and $K$, $E$ correspond to the complete elliptic integral functions of the first and second kind respectively (see Ref.~\citep{abramowitz1964}, p. 590). The first term in Eq~\eqref{eq:potential_nff} is the flag's contribution and $\phi_{{\cal W}_i}$ is the contribution of the wake of non-dimensional size $L_w$. Equation~\eqref{eq:potential_nff} will allow us to study the influence of the wake on the flow reconstruction in the following paragraph by means of a simple wake model. Later on, an additional assumption termed \textit{far-field} approximation will be introduced for conveniency, and its range of validity will also be examined.  
}
\begin{figure}
\centering
\begin{tabular}{cc}
\includegraphics[width=6.5cm]{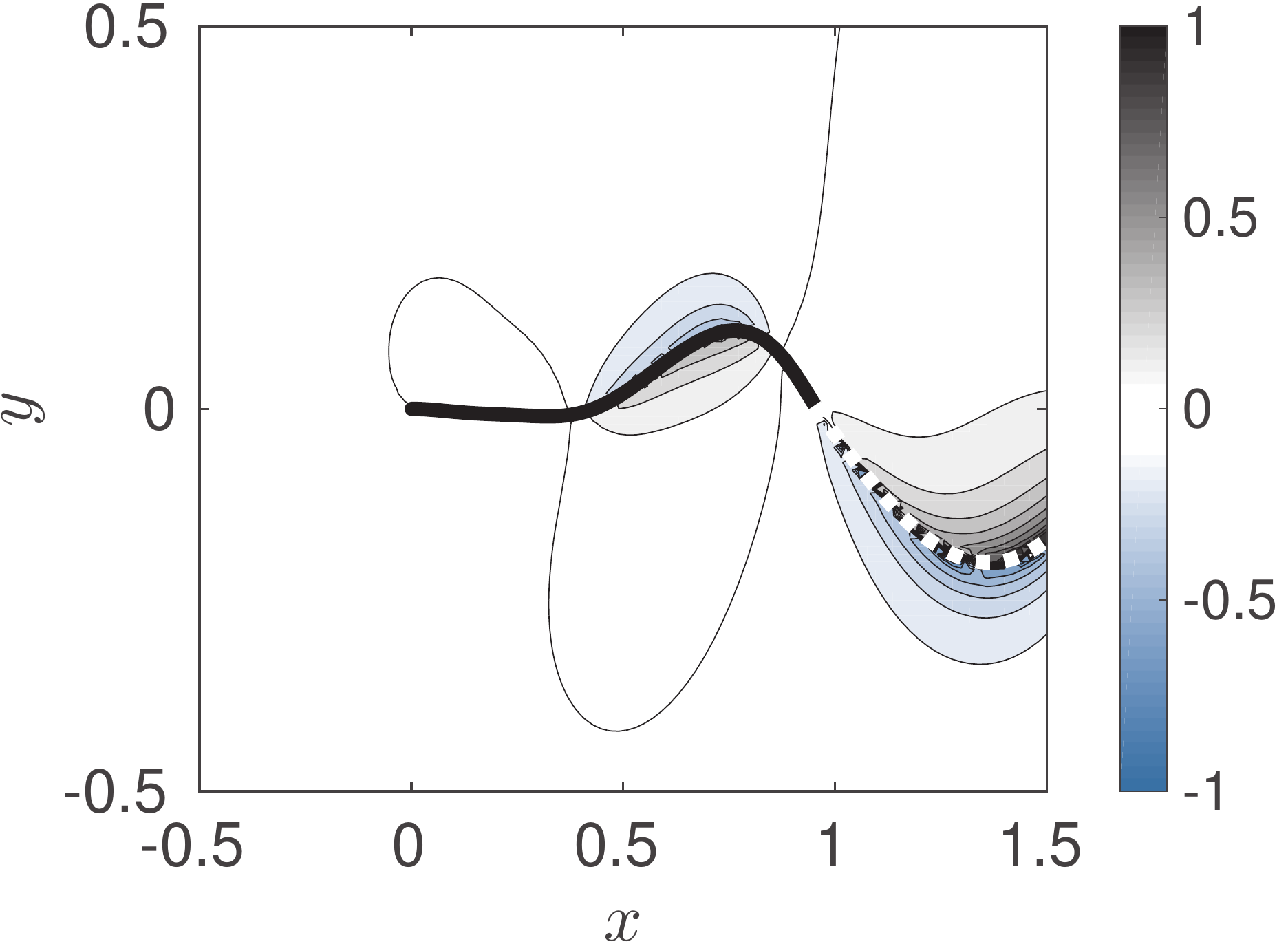} & 
\includegraphics[width=6.5cm]{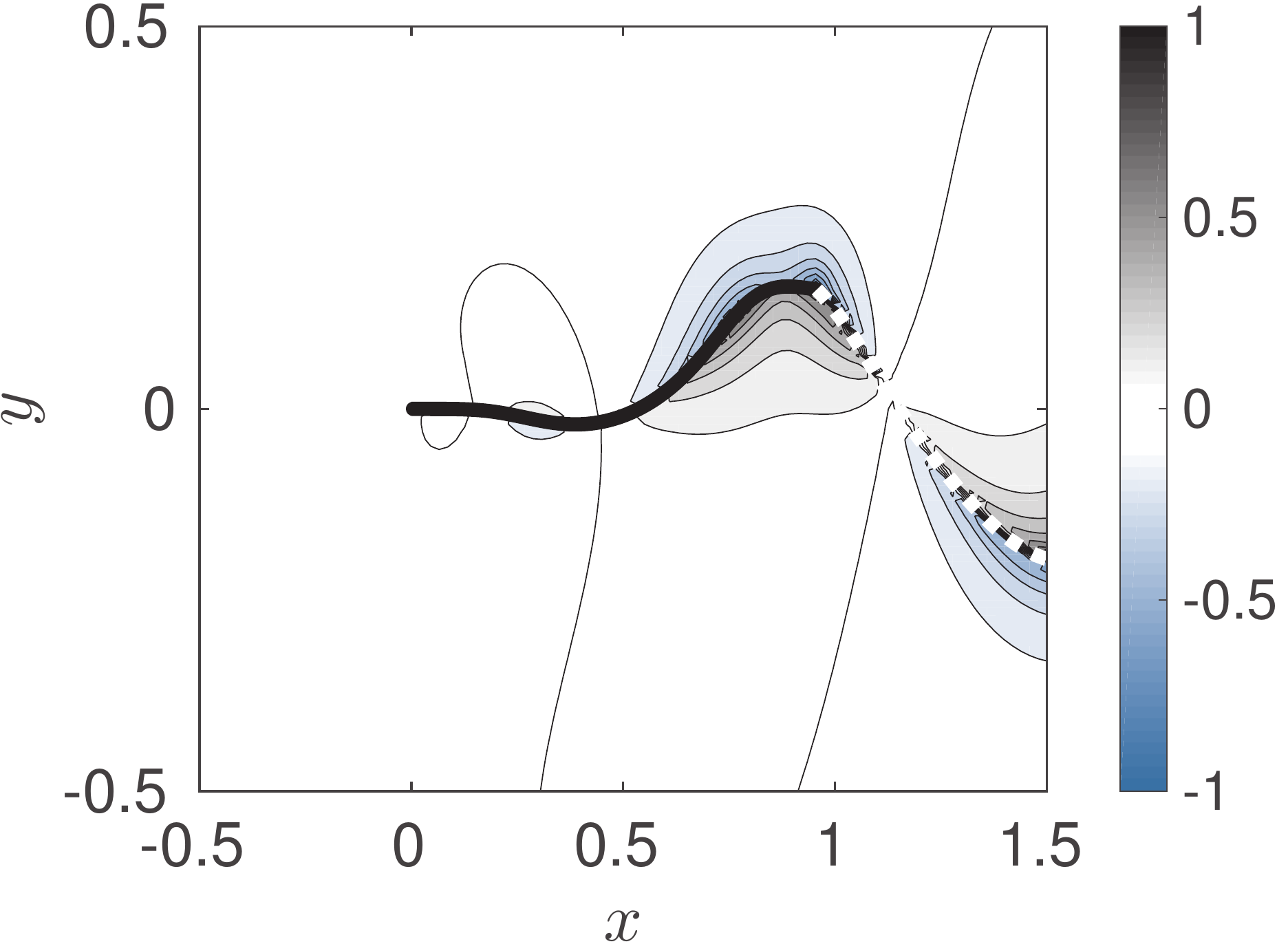} \\
$(a)$ $t$ & $(b)$ $t+T/8$\\
\includegraphics[width=6.5cm]{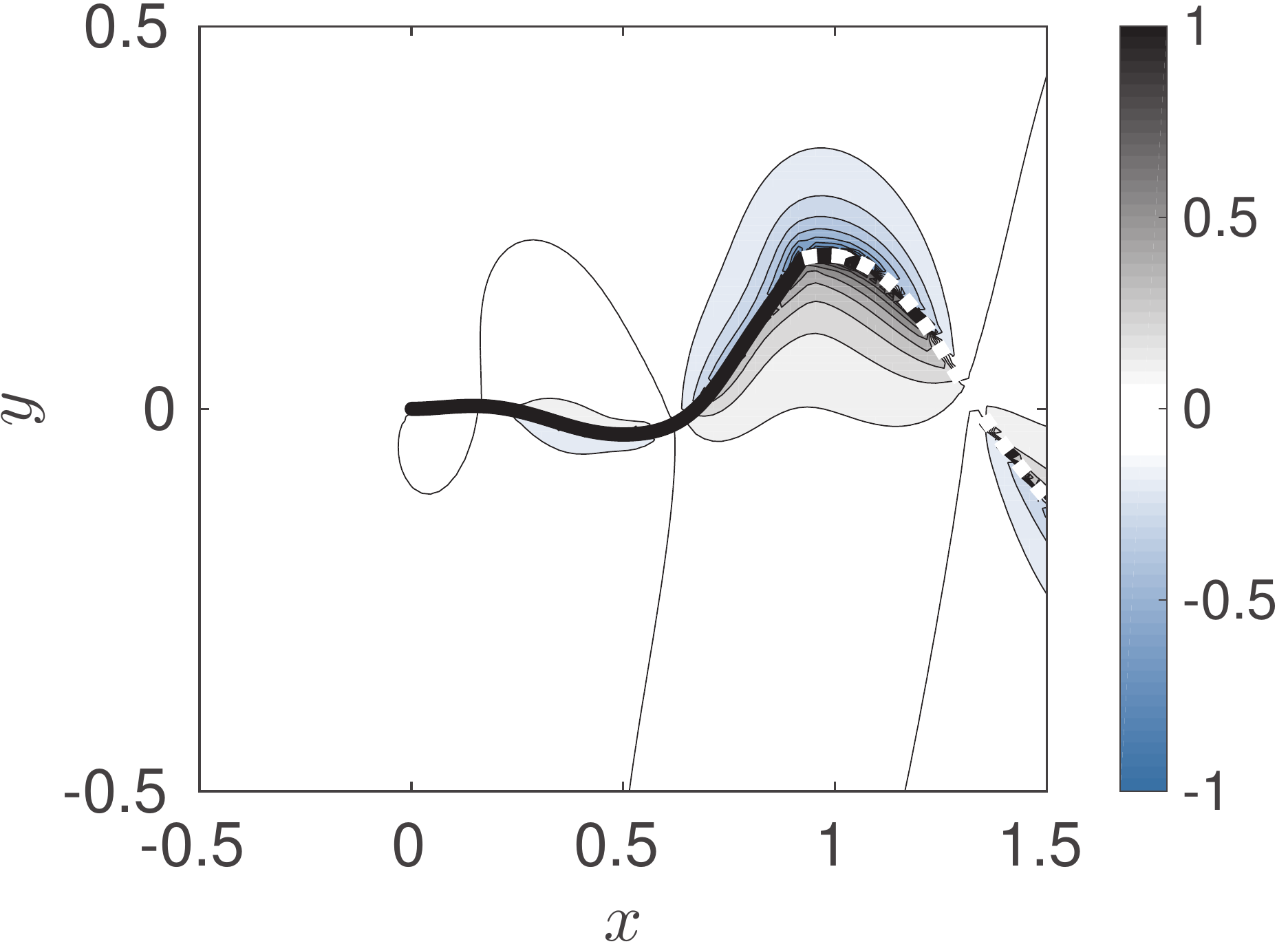} & 
\includegraphics[width=6.5cm]{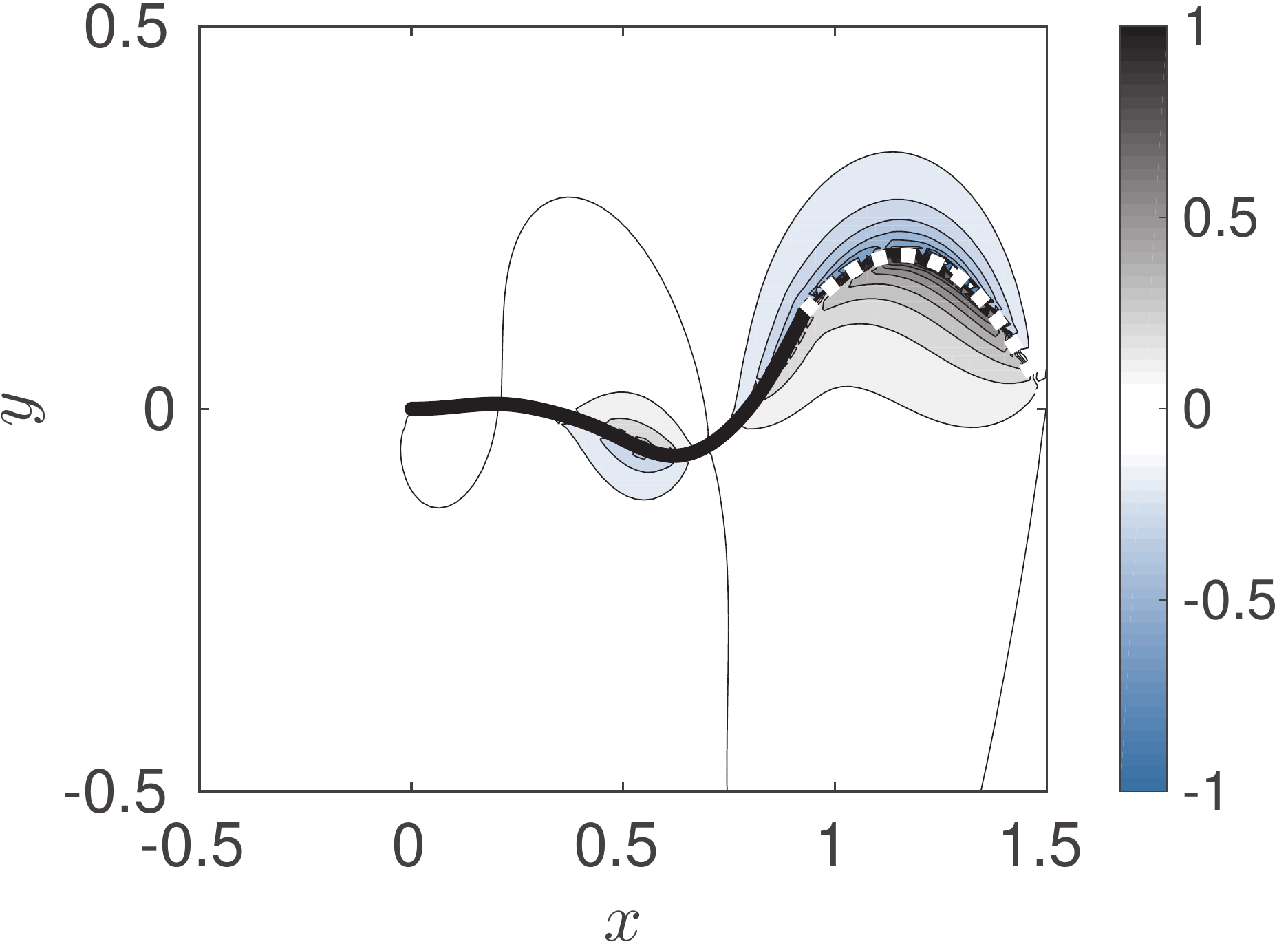} \\
$(c)$ $t+2T/8$ & $(d)$ $t+3T/8$
\end{tabular}
\caption{\label{fig:flow_LAEBT_contours} Flow created in the midplane $z=0$ during a flapping sequence for $M^*=10$, $U^*=20$ and $H^*=0.1$. 
Instantaneous normalized velocity potential obtain from Eqs~\eqref{eq:representation_th} and \eqref{eq:pot_jump} using a \textit{frozen} wake model of non-dimensional length $L_w=2$. The flag's (resp. wake's) position is shown by a thick black (resp. dashed) line.}
\end{figure}

\change{\subsubsection{Role of the wake}}
Solving Eqs~\eqref{eq:potential_nff}, the flow around the flag is obtained everywhere provided the wake characteristics \change{(position and circulation)} are known. 
In order to assess the effect of the wake on the created flow, a simplified wake model is constructed. We neglect auto-induction and therefore assume that the vorticity shed at the trailing edge is only advected downstream by the uniform flow. This wake model is referred to as \textit{frozen wake} in the following and has already been considered in previous studies on flexible bodies~\citep{candelier2011}.  Figure \ref{fig:flow_LAEBT_contours} shows an example of velocity potential contours obtained from equation Eqs~\eqref{eq:potential_nff} with such a simplified wake model. From this flapping sequence, it can be seen that the flow varies in the stream-wise direction on length scales of the order of the flapping wave number, that is of the order of $L$ for the first flapping modes. 

\begin{figure}
\centering
\begin{tabular}{cc}
\includegraphics[width=6.5cm]{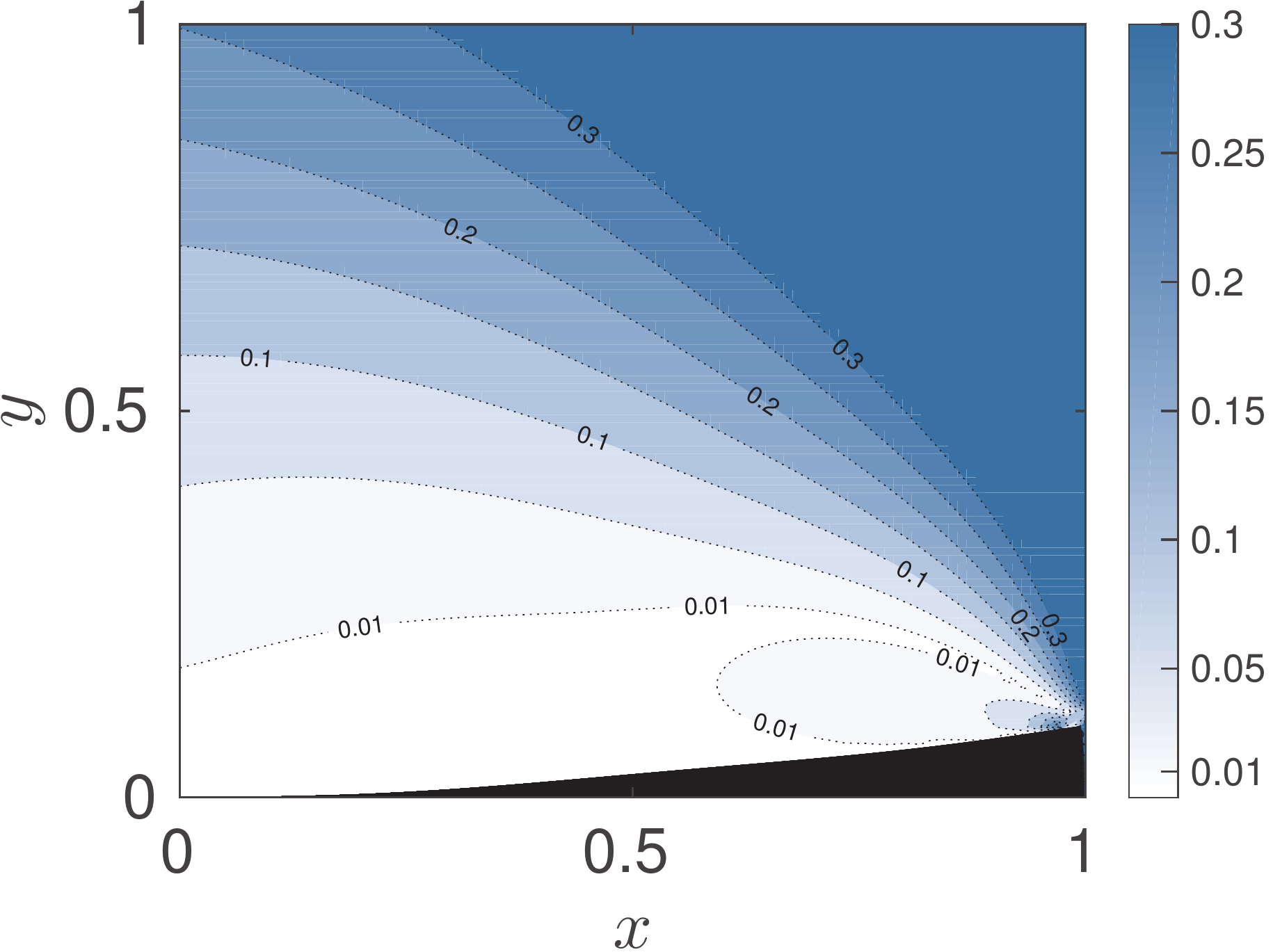}& \hspace{0.5cm}
\includegraphics[width=6.5cm]{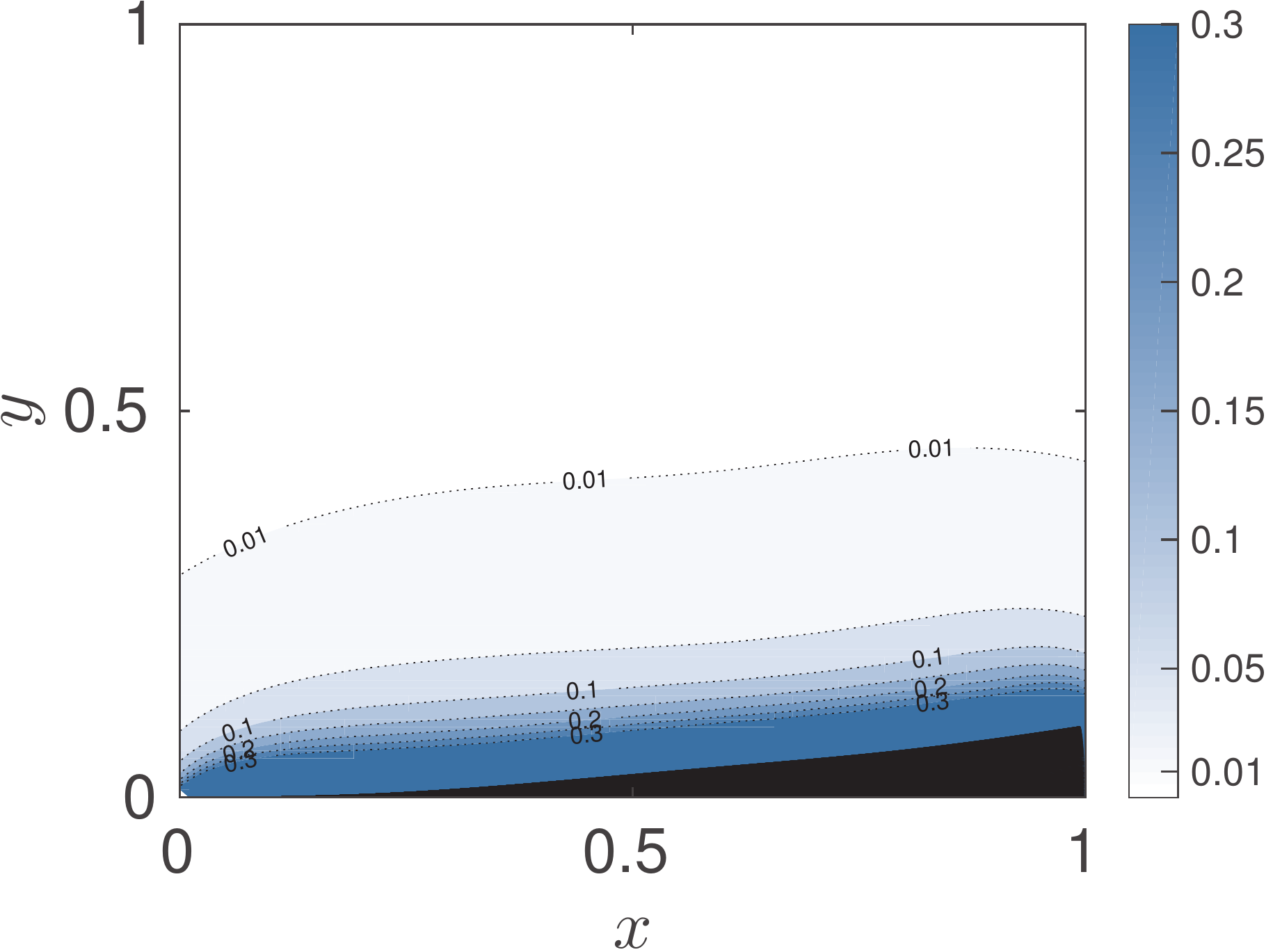}\\
$(a)$  & $(b)$ 
\end{tabular}
\caption{\label{fig:error_maps} Relative error maps for the flow reconstruction in the midplane $z=0$ in the case $M^*=3$, $U^*=15$ and $H^*=0.1$. $(a)$ Error committed when the wake is neglected. The error is defined as $|[rms(v_{L_w=2}) - rms(v_{L_w=0})]/rms(v_{L_w=2})|$ where the case without wake is compared to the case with a \textit{frozen} wake of non-dimensional length $L_w=2$.  $(b)$ Error due to far-field approximation ($FF$) defined as $|[rms(v_{FF}) - rms(v)]/rms(v_{FF})|$. Wakes are neglected in this case. The black area delineates the envelope of the flapping flag.} 
\end{figure}

\change{The effect of the wake on the created flow field is now investigated in figure \ref{fig:error_maps}$(a)$ which corresponds to the relative error map obtained by comparison between results with a wake of non-dimensional length $L_w=2$ and results obtained without taking the wake into account. The white area corresponds to locations where the error associated with ignoring the wake is less than $1\%$ while the darkest blue region indicates an error larger than $30\%$. For this value of $H^*$, the length of the wake does not significantly change the result provided $L_w\geq 1$ (not shown). In these cases, we therefore obtain that the wake influence is weak while $y< 0.3$ (for $H^*=0.1$) and far enough from the trailing edge as the error becomes important for distances ${\cal O}(H^*)$ from this location. 
When considering two flags side by side, it is reasonable to neglect wake effect in the coupling terms within the range $d^* < 0.3$. Even though the flow is not well predicted close to the trailing edge, it will be verified that it does not significantly affect the global dynamics whose prediction is the main focus of the present article.}

\change{\subsubsection{Far-field approximation (FF)}}
\change{Equation~\eqref{eq:potential_nff} can be reformulated when looking at the flow field away at distances from the flag that are large compared to $H^*$. Using the asymptotic expansion of $E(X)$ and $K(X)$ for $X\ll 1$, the following approximation of the flow potential is obtained in the midplane $z=0$,
\begin{equation}
\label{eq:potential_ff}
\phi_{i}(\boldsymbol{x},t)= -\frac{{H*}^2}{16}\int_{0}^{1}{ \frac{\tilde{u}_{n_i}(\boldsymbol{x}-\boldsymbol{x}_i) \cdot\boldsymbol{e}_{n_i}}{[(x-x_i)^2 + (y-y_i)^2]^{3/2}}ds_j(\boldsymbol{x}_i)},
\end{equation}
if $\phi_{{\cal W}_i}$ is further neglected. This approximation is equivalent to assuming that the Green function is independent of the spanwise coordinate in Eq.~\eqref{eq:representation_th}.
The range of validity of the far-field approximation can be estimated from Figure \ref{fig:error_maps}$(b)$ where the error shows that this approximation essentially affects the flow for lateral distances to the flag of order $H^*$. Equation.~\eqref{eq:potential_ff} therefore provides a simplified version of Eq.~\eqref{eq:potential_nff} which is valid far from the flag (compared to $H^*$) and in regions where wakes do not contribute significantly to the flow; it is therefore valid for intermediate lateral distances ($0.1<y<0.3$ in the present case where $H^*=0.1$). This form of the velocity potential given by Eq.~\eqref{eq:potential_nff} is introduced here only as a matter of convenience, as it allows for faster simulations and simplifies the analysis; but the present method could be applied directly using Eq.~\eqref{eq:potential_nff}. In addition, as detailed in the following, far-field approximation is consistent with further modeling steps for two flags and has little impact on the global dynamic when $d^*>H^*$. }\\

\subsection{LAEBT in a weakly non-uniform potential flow}
\label{subsec:LAEBT_NU}
Lighthill's LAEBT was recently generalized by Ref.~\cite{candelier2013} to the case of a slender structure in a weakly non-uniform potential flow. In the classical LAEBT, the uniform incoming flow and the flow created by the flapping flag can respectively be termed \textit{ambient flow} and \textit{perturbed flow}. 
Ref.~\cite{candelier2013} extends LAEBT for weakly non-uniform ambient flows, \emph{i.e.} for cases where the ambient flow varies on length scales much larger than the cross-section \change{dimensions}.

\change{Under this assumption, the local ambient flow can be expended in Taylor series in each cross section around the center line of the body.} This provides a decomposition of the perturbed flow which accounts for non-uniformities of the ambient flow. Such a procedure eventually leads to an expression for the fluid force exerted on a body immersed in a weakly non-uniform and potential flow, which has been successfully implemented to simulate swimming of a slender fish in a Von Karman vortex street~\citep{candelier2013}. 

\change{
In this paragraph, we present a brief summary of this result, and use dimensional quantities (in capital letters) to clarify the physical origin of the different contributions to the force. The reader is referred to the original study of Ref.~\cite{candelier2013} for more details. Considering a slender structure with center-line position $\boldsymbol{X}_{0}$ immersed in a potential and weakly non-uniform ambient flow with velocity $\boldsymbol{V}(\boldsymbol{X},T)$ and pressure $P(\boldsymbol{X},T)$, body-fitted coordinates $X^n$ and $X^{\tau}$ are introduced and respectively correspond to normal and tangential positions. For planar motions of the structure, the dimensional local pressure force exerted on the solid is obtained as (see Ref.~\citep{candelier2013}, Eq.~(4.5))
\begin{equation}
\label{eq:lighthill_non_uniform_candelier}
{\boldsymbol{F}} =  - S \left.\nabla P\right|_{\boldsymbol{X}=\boldsymbol{X}_{0}} -  \left[\frac{\partial  M_a U_{n}  {\boldsymbol{e}_{n}}}{\partial T} -\frac{\partial  M_a U_{n} U_{\tau}  {\boldsymbol{e}_{n}}}{\partial X^{\tau}}  +\frac{1}{2} \frac{\partial M_a U_{n}^2 {\boldsymbol{e}_{\tau}}}{\partial X^{\tau}} \right] - M_a U_{n}  \left. \frac{\partial \boldsymbol{V}}{\partial  X^n} \right|_{\boldsymbol{X}=\boldsymbol{X}_{0}}  \,
\end{equation}
 with $S$  the surface area of the body's local cross-section, and $M_a $ the added mass associated to its normal displacement. 
 Note that the body shape (and therefore $S$ and $M_a$) may slowly vary along the tangential direction in this formalism.
 In addition, $U_n$ and $U_{\tau}$ correspond to the components of the local relative velocity between the solid and the ambient flow which are defined as
\begin{equation}
\label{eq:relative_velocity_dim}
U_{n}\boldsymbol{e}_{n}+U_{\tau}\boldsymbol{e}_{\tau}=\frac{\partial \boldsymbol{X}_{0}}{\partial T} - \boldsymbol{V}(\boldsymbol{X}=\boldsymbol{X}_{0}).
\end{equation} 
The physical origin of the three terms in Eq.~\eqref{eq:lighthill_non_uniform_candelier} can be understood as follows:
\begin{itemize}\renewcommand{\labelitemi}{$-$}
\item The first term is due to non-uniformities of the ambient pressure and can physically be interpreted as a generalization of Archimedes force which vanishes in the case of the infinitely-thin flag considered here ($S\rightarrow 0$).
\item The second term corresponds to the classical LAEBT expression \cite{singh2012} in which the relative velocity defined by Eq.~\eqref{eq:relative_velocity_dim} now takes into account the non-uniformities of the ambient flow.
\item The third term is an additional contribution due to structure's motion within the ambient velocity gradient.
\end{itemize}
}

\change{
\subsection{LAEBT in a weakly non-uniform potential flow: application to flag geometry}
Moving back to the non-dimensional framework introduced in \S \ref{sec:model} the above theory is now applied to the specific flag geometry investigated in this work.
Considering a flag ${\cal F}_i$ placed in a weakly non-uniform and potential ambient flow $\boldsymbol{v}$, the general expression of the reactive fluid force shown in Eq.~\eqref{eq:lighthill_non_uniform_candelier} simplifies for an infinitely-thin flag with uniform added-mass coefficient ($m_a=\pi/4$). Furthermore, applying the inextensibility of the structure and irrotationality of the ambient flow, the tangential component of the force in Eq.~\eqref{eq:lighthill_non_uniform_candelier} can be shown to vanish exactly. If the ambient flow is further symmetric with respect to the mid-plane, the local reactive force exerted on ${\cal F}_i$ is purely normal, and its non-dimensional form can be written as
}
\begin{equation}
\label{eq:lighthill_non_uniform}
{\boldsymbol{f}_\textrm{react}}_i = - m_a H^* M^* \left( \frac{\partial u_{n_i}}{\partial t} - \frac{\partial u_{n_i} u_{\tau_i}}{\partial s_i}  + \frac{u_{n_i}^2}{2} \frac{\partial \theta_i }{\partial s_i} +  u_{n_i} \boldsymbol{e}_{n_i}.\left[ \left. \nabla \boldsymbol{v} \right|_{\boldsymbol{x}_i=\boldsymbol{x}_{i0}}  \right].{\boldsymbol{e}_{n_i}}  \ \right) {\boldsymbol{e}_{n_i}}
\end{equation}
with relative velocity
\begin{equation}
\label{eq:relative_velocity}
u_{n_i}\boldsymbol{e}_{n_i}+u_{\tau_i}\boldsymbol{e}_{\tau_i}=\frac{\partial \boldsymbol{x}_{i0}}{\partial t} - \boldsymbol{v}(\boldsymbol{x}_i=\boldsymbol{x}_{i0}).
\end{equation} 
Equation~\eqref{eq:lighthill_non_uniform} generalizes LAEBT to the motion of a flag in non-uniform flow. Inhomogeneities of the ambient flow appear explicitly in the last term and implicitly in the others through the relative velocity defined in Eq.~\eqref{eq:relative_velocity}. The Elongated Body Theory (EBT) can be extended to non-uniform flows in a similar fashion, and corresponds to the leading order expansion of the previous equation in the limit of small displacements. Since $\boldsymbol v$ is only weakly non-uniform, the last term in Eq.~\eqref{eq:lighthill_non_uniform} is quadratic in the small flapping amplitude regime and should be discarded: the force expression in the EBT is therefore formally identical in uniform and weakly non-uniform flows, and only differ in the definition of the appropriate relative velocity. 

Additionally, the reactive contribution given by Eq.~\eqref{eq:lighthill_non_uniform} from LAEBT must be complemented by a resistive contribution which should also be modified to account for flow non-uniformities. 
In line with Ref.~\cite{eloy2012b} skin drag is neglected and we model the drag associated with lateral flow separation as
\begin{equation}
\label{eq:f_resist}
{\boldsymbol{f}_\textrm{resist}}_i = -\frac{1}{2} M^* C_d u_{n_i} |u_{n_i}|  \boldsymbol{e}_{n_i},
\end{equation}
with $C_d=1.8$ for a flat plate and $u_{n_i}$ defined in Eq.~\eqref{eq:relative_velocity}.\\

The combination of ${\boldsymbol{f}_\textrm{react}}_i$ and ${\boldsymbol{f}_\textrm{resist}}_i$ finally provides a model for the local fluid force applied on a slender flag immersed in the weakly non-uniform flow $\boldsymbol{v}$. \change{The weak non-uniformity means that the components of $\boldsymbol{v}$ are not significantly varying over $\mathcal{O}(H^*)$ length scales. }In particular, this explains why only the flow at the center line is needed to obtain the fluid forces in Eqs~\eqref{eq:lighthill_non_uniform} and \eqref{eq:f_resist}.
In the following these expressions are used to model the configuration of two slender flags by considering that the ambient flow corresponds to the superposition of the uniform axial flow and the flow created by the neighbouring flag.

\subsection{LAEBT for two slender flags}
\change{For a specific intermediate range of non-dimensional distances $d^*$, combination of the results of the two previous sections provides an extension of LAEBT to the case of two slender and infinitely thin structures placed side by side.} The underlying idea is to consider that each flag is flapping in the non-uniform flow corresponding to the superposition of the uniform incoming flow and the flow created by its neighbour. From \S \ref{subsec:LAEBT_NU} the local fluid forces exerted on each flag can therefore be modeled as ${{\boldsymbol{f}}_\textrm{fluid}}_i= {{\boldsymbol{f}}_\textrm{react}}_i +  {{\boldsymbol{f}}_\textrm{resist}}_i $, where $ {{\boldsymbol{f}}_\textrm{react}}_i$ and $ {{\boldsymbol{f}}_\textrm{resist}}_i$ are given by Eqs~\eqref{eq:lighthill_non_uniform} and \eqref{eq:f_resist} respectively, and in which the velocity $\boldsymbol{v}$ must be replaced by $\boldsymbol{e}_x + \nabla \phi_j$ where $\phi_j$ corresponds to the velocity potential created by ${\cal F}_j$ (with $j \ne i$) and obtained from Eq.~\eqref{eq:potential_nff}.  

\change{Based on the conclusions of  \S \ref{subsec:flow_LAEBT} and unless otherwise stated, the influence of the wakes is neglected in the following which focuses on regimes where the flags are close compared to their length. Doing so effectively overlooks the modification of the flow field in the trailing edge's immediate vicinity, but we show in the following that this assumption has essentially no effect on the overall dynamics.} 

In addition, the LAEBT extension leading to the reactive force is only valid in the case of a weakly non-uniform flow, i.e. if the components of $\boldsymbol{v}$ are not varying much on length scales of the order of $H^*$. The present fluid model is therefore valid if all parts of the flags remain far compared to $H^*$, a condition which corresponds to $H^*\ll d^*$ \change{for small amplitudes. Large amplitude cases require more care as it depends on the synchronization phase between flags, but it is worth noting that it will automatically hold in the case of in-phase motion for which the distance between flags does not get significantly smaller than $d^*$.}
This weakly non-uniform restriction legitimates the far-field approximation and $\phi_j$ is therefore calculated using Eq.~\eqref{eq:potential_ff}. 

As a conclusion, Equations~\eqref{eq:momentum}--\eqref{eq:inextensibility} for both flags' dynamics coupled to the fluid model provided by Eq.~\eqref{eq:lighthill_non_uniform}--\eqref{eq:f_resist} with $\boldsymbol{v} = \boldsymbol{e}_x + \nabla \phi_j$ and $\phi_j$ given by Eq.~\eqref{eq:potential_ff} provide a model for two side by side flags with geometric parameters in the range $H^*\ll d^*\ll 1$.

\section{Linear case}
\label{sec:linear}
For small lateral displacements $y_i(s_i,t)$ of flag ${\cal F}_i$ ($i=1,2$), Eqs~\eqref{eq:momentum}--\eqref{eq:inextensibility} and \eqref{eq:lighthill_non_uniform}--\eqref{eq:f_resist} can be linearized around the equilibrium position, $y_i(s,t)=0$, leading to the EBT formulation of the two-flag problem:

\begin{figure}
\centering
\begin{tabular}{cc}
\includegraphics[width=6.5cm,trim = 1cm 1cm 1.5cm 0cm, clip]{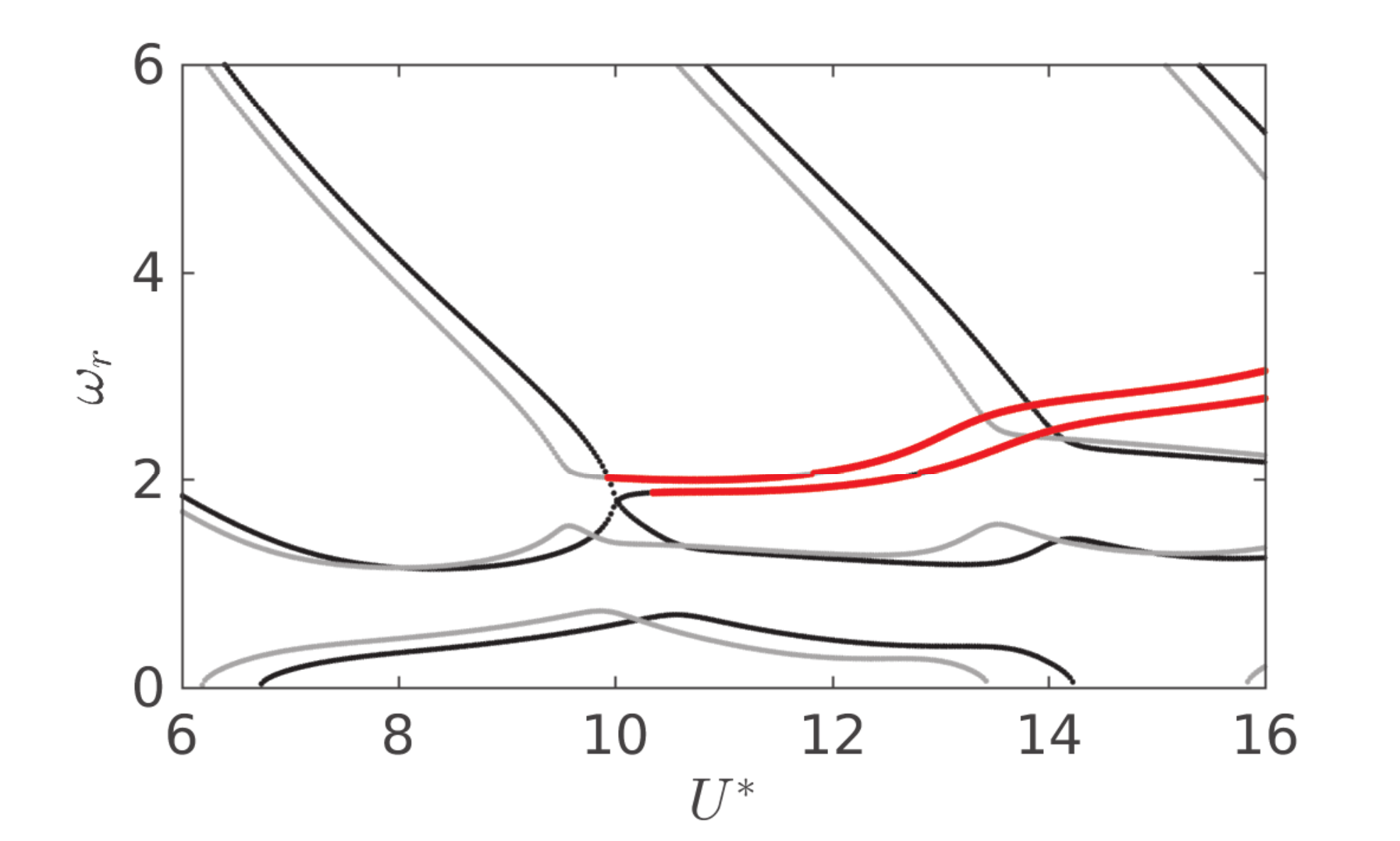}&\includegraphics[width=6.5cm,trim = 1cm 1cm 1.5cm 0cm, clip]{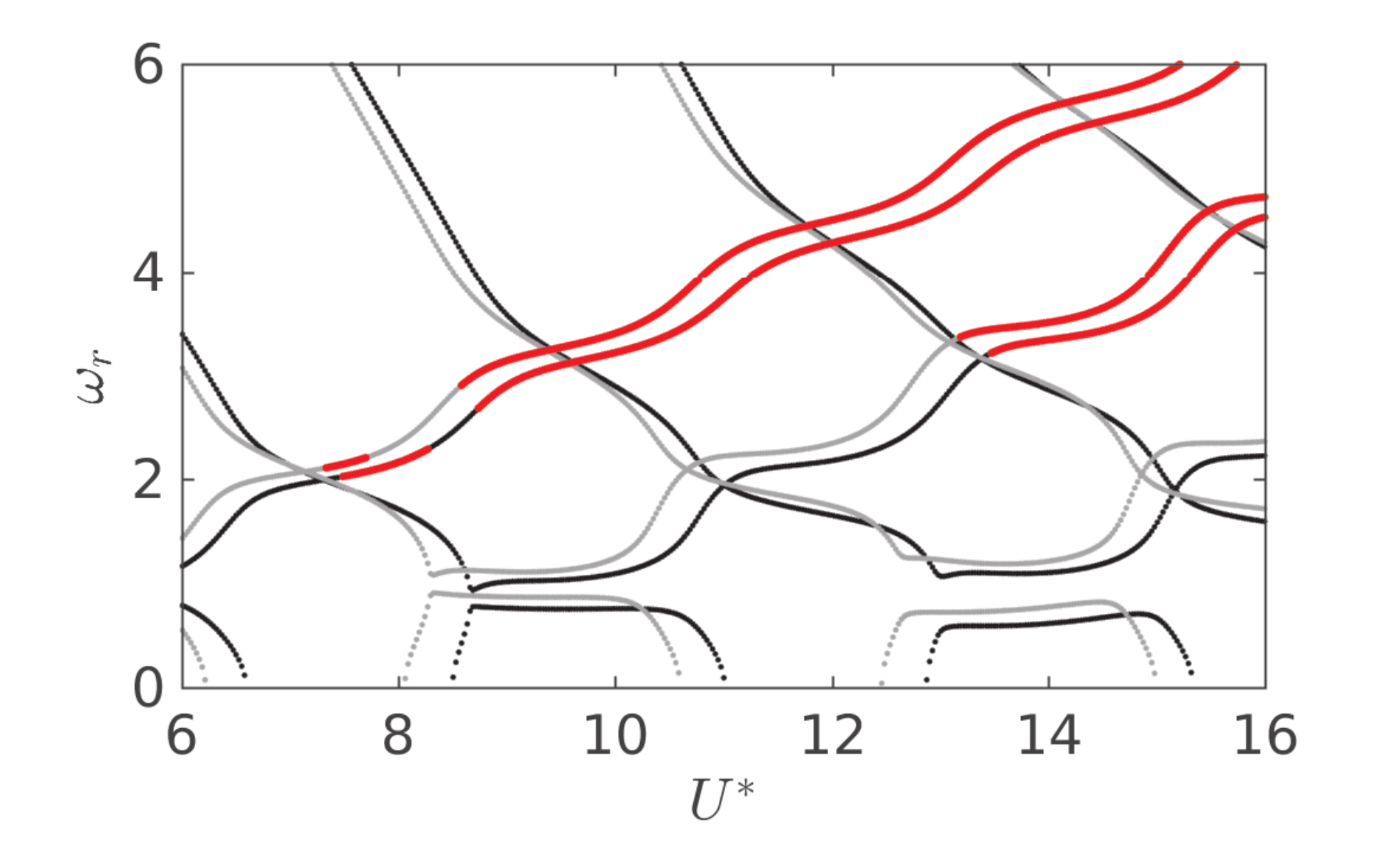}\\
\includegraphics[width=6.5cm,trim = 1cm 0cm 1.5cm 0cm, clip]{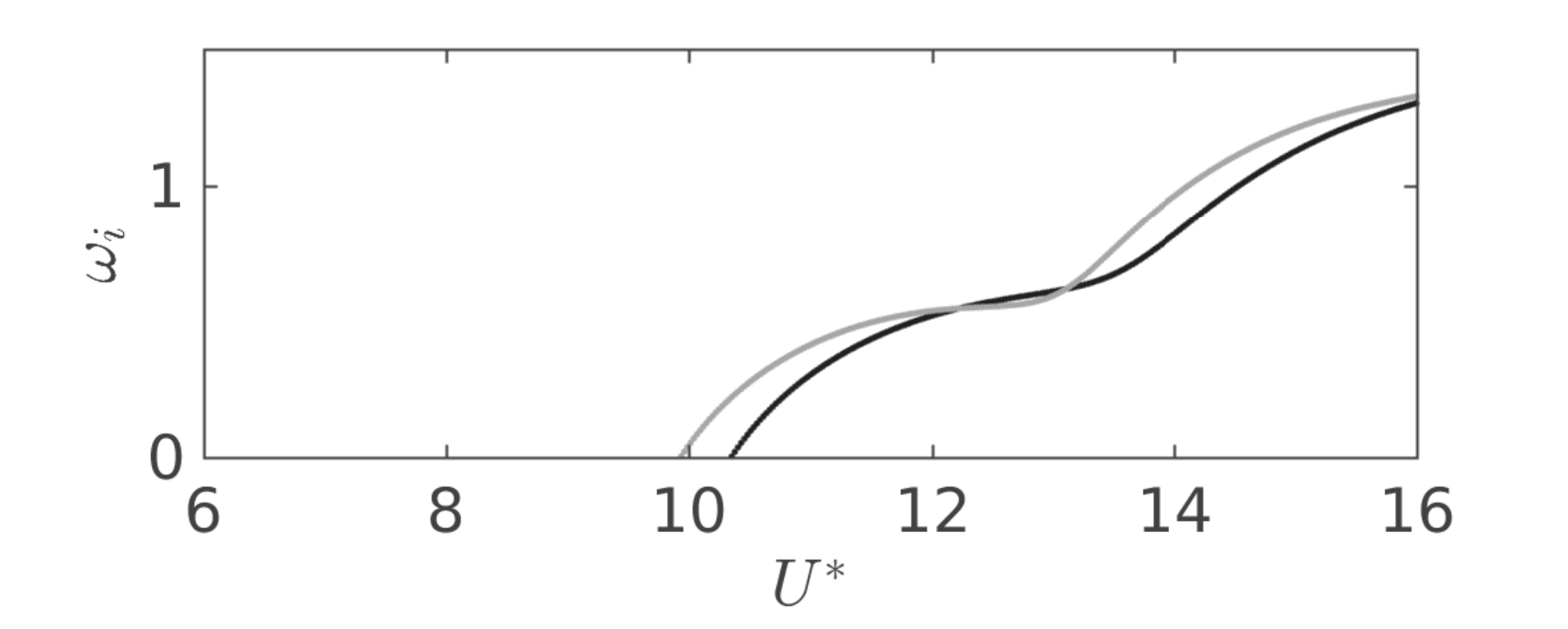}&\includegraphics[width=6.5cm,trim = 1cm 0cm 1.5cm 0cm, clip]{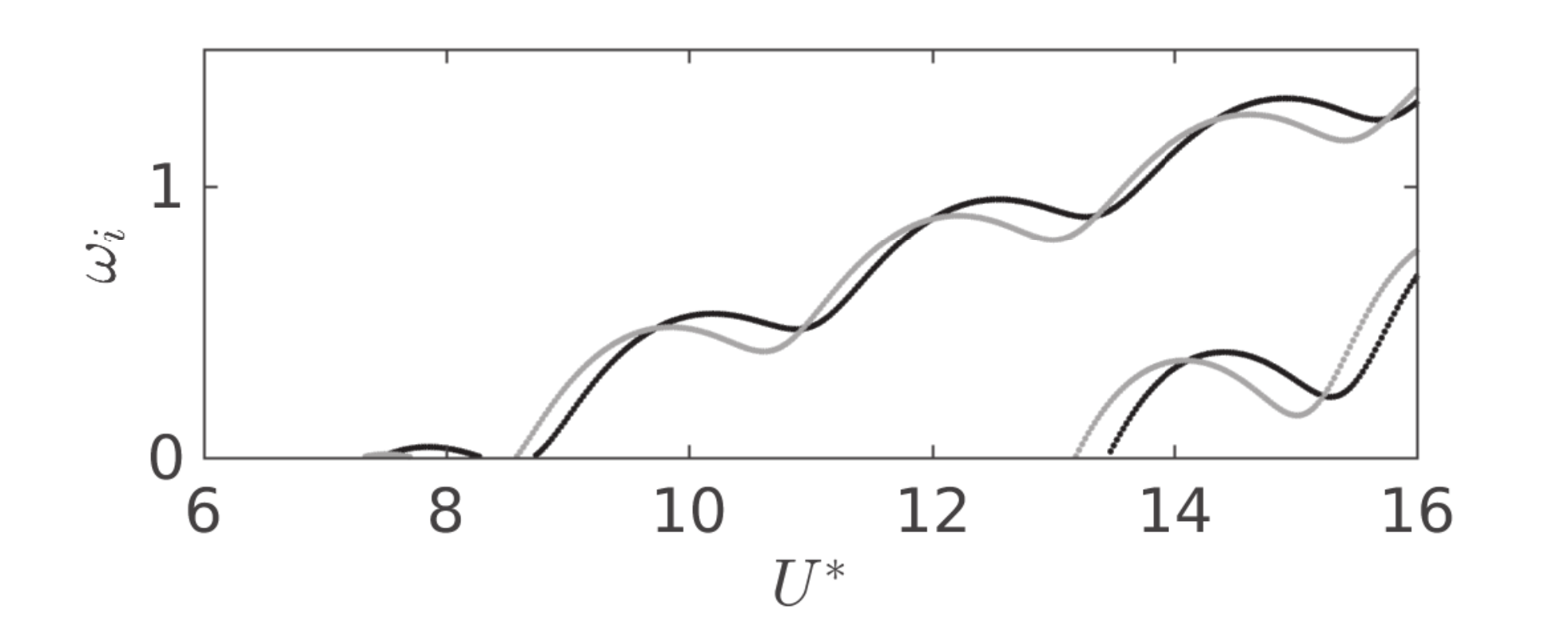}\\
(a) $M^*=10$ & (b) $M^*=27$
\end{tabular}
\caption{\label{fig:global_freq} Frequencies (top) and growth rates (bottom) for $H^*=0.1$ and $d^*=0.15$. The black lines (resp. gray lines) correspond to in-phase (reps. out-of-phase) modes. Frequencies corresponding to positive growth rates are shown in bold red.}
\end{figure}

\begin{align}
\label{eq:coupled_linear}
 \frac{\partial^2 y_i}{\partial t^2}+ \frac{1}{{U^*}^2}  \frac{\partial^4 y_i}{\partial s_i^4}  &=-m_a H^* M^*\left(\frac{\partial^2 y_i}{\partial t^2} + 2 \frac{\partial^2 y_i}{\partial t \partial s_i} +  \frac{\partial^2 y_i}{\partial s_i^2} \right) + m_a H^* M^* \left(\frac{\partial }{\partial t} + \frac{\partial }{\partial s_i} \right)v_{ji}, 
\end{align}
\change{with $v_{ji}$ the linear lateral velocity induced by ${\cal F}_j$ ($j\neq i$) on ${\cal F}_i$ and is obtained from Eq.~\eqref{eq:potential_ff}} 
\begin{equation}
\label{eq:lin_vji}
v_{ji}(s_i)  = \frac{{H^*}^2}{16} \int_0^1{  \left( \frac{\partial y_{j}}{\partial t} +  \frac{\partial y_{j}}{\partial s_{j}}  \right)  \frac{2 {d^*}^2 - (s_i-s_j)^2}{[{d^*}^2 + (s_i-s_j)^2]^{5/2}} ds_j}.
\end{equation}

\change{
Equations~\eqref{eq:coupled_linear}--\eqref{eq:lin_vji} provide a simplified linear system for $y_1(s,t)$ and $y_2(s,t)$, which also provides important insight on the different fluid contributions (right-hand side of Eq.~\eqref{eq:coupled_linear}) and justifies \textit{a posteriori} some modeling assumptions.}
This discussion will be made using a classification of the fluid terms in powers of $H^*$ and $d^*$.

\change{
The first fluid term of Eq.\eqref{eq:coupled_linear} corresponds to the traditional EBT contribution \citep{eloy2007}, while  the second results from the coupling with the neighbouring flag.  Both fluid terms correspond to the flow forcing over the local cross-section, leading to a pre-factor $H^*$ with the present set of non-dimensional numbers. For the coupling term, a span-wise integration is performed on the forcing flag leading to the factor ${H^*}^2$ in the induced velocity.
The integral in Eq.~\eqref{eq:lin_vji} scales as $1/d^{*2}$, and the coupling terms therefore finally scales as ${\cal O}({H^*}^3/{d^*}^2)$. 
We therefore obtain that EBT and coupling terms scale respectively as ${\cal O}({H^*})$ and  ${\cal O}({H^*}^3/{d^*}^2)$. 
In order to validate the model, these terms should be compared to the first correction of EBT for a single flag (which includes the first order contribution of the flag own wake), which scales as ${\cal O}({H^*}^3 \ln H^*)$ \citep{eloy2010}. For a flag aspect ratio $H^*$, the coupling term decreases with $d^*$ and becomes as small as the EBT correction at some point. For $H^*=0.1$ for instance, it is found that the coupling term has the same magnitude as the EBT correction for $d^*\approx 0.7$.  For distances around and larger than this value, it would therefore be inconsistent to retain hydrodynamic coupling terms while neglecting EBT corrections. 
The present approach therefore only works while $d^*<<1$. In order to extend this approach for larger values of $d^*$, EBT corrections (and therefore the contribution of the flag's own wake) should be included and the coupling terms due to the neighbour's wake should also be taken into account. }

\change{
In addition, it should be noted that higher order coupling terms obtained by taking into account, during the flow reconstruction, the flag's immersion in a non-uniform flow (i.e. replacing $\tilde{u}_{n_i}$ by $u_{n_i}$ in Eq.~\eqref{eq:pot_jump}) scale as ${\cal O}({H^*}^5/{d^*}^4)$. This term is therefore  neglected in the present study. }\\

Searching for $y_i(s_i,t) = Y_i(s_i)e^{-i \omega t}$ with clamped-free boundary conditions and using spatial Chebyshev collocation, the system \eqref{eq:coupled_linear}--\eqref{eq:lin_vji} is rewritten as a generalized eigenvalue problem for $\omega$ and $[Y_1(s_1), Y_2(s_2)]$. 
The real and imaginary parts of $\omega$  are  the frequency $\omega_r$ and growth rate $\omega_i$ of the corresponding mode, respectively, and $\omega_i>0$ denotes instability. The problem's symmetry imposes that $Y_1=\pm Y_2$, i.e. flags are either in-phase or out-of-phase~\citep{michelin2009}. 

\begin{figure}
\centering
\includegraphics[width=0.55\textwidth]{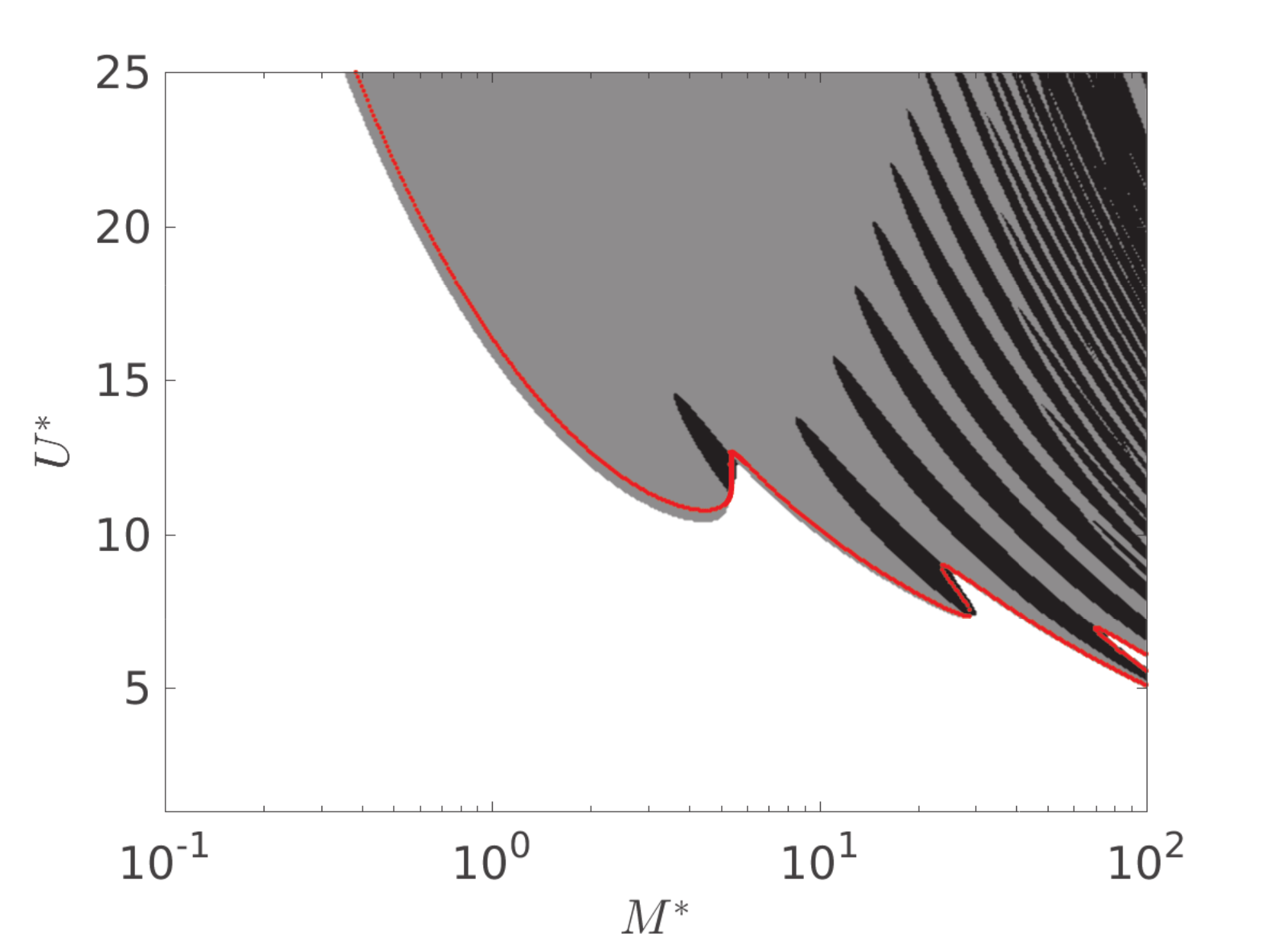} 
\caption{\label{fig:map_M} Synchronization in the most unstable mode for $H^*=0.1$ and $d^*=0.15$. In-phase (resp. out-of-phase) dominant modes correspond to black (resp. grey) regions. White regions correspond to a stable configuration. The red line indicates the stability threshold of a single flag ($d\rightarrow\infty$).}
\end{figure}

\change{Figure~\ref{fig:global_freq} shows that the two-flag configuration remains stable up to a critical reduced velocity $U^*_c$ when the out-of-phase mode becomes unstable for the  values of $M^*$, $H^*$ and $d^*$ considered}. Increasing  $U^*$ further, successive switches are observed  in the synchronization of the most unstable mode (out-of-phase/in-phase). The map of the most unstable mode in the parameter space $(M^*, U^*)$ is shown on Figure \ref{fig:map_M} and reveals that such switching phenomena are found for large values of $M^*$. From this general picture, we conclude that in-phase modes are predominantly expected for large $M^*$ and out-of-phase modes for small  $M^*$, a trend reminiscent of the linear predictions in the two-dimensional limit~\citep{michelin2009}. In addition, the presence of a neighbour has a destabilizing effect for most $M^*$ as evidenced on Figure~\ref{fig:map_M} by comparison with the results for a single flag. 

\begin{figure}
\centering
\begin{tabular}{cc}
\includegraphics[width=0.45\textwidth]{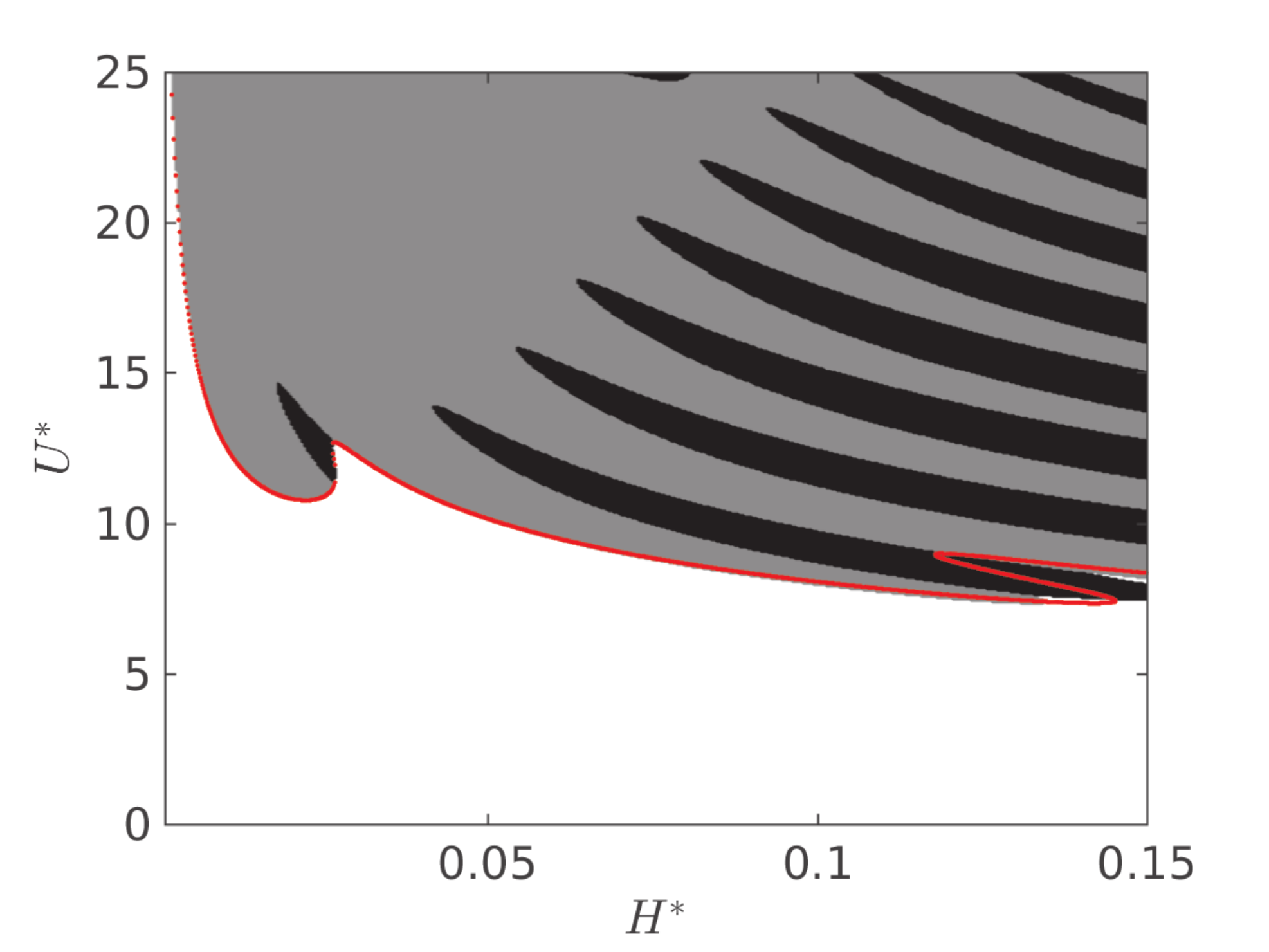} & \includegraphics[width=0.45\textwidth]{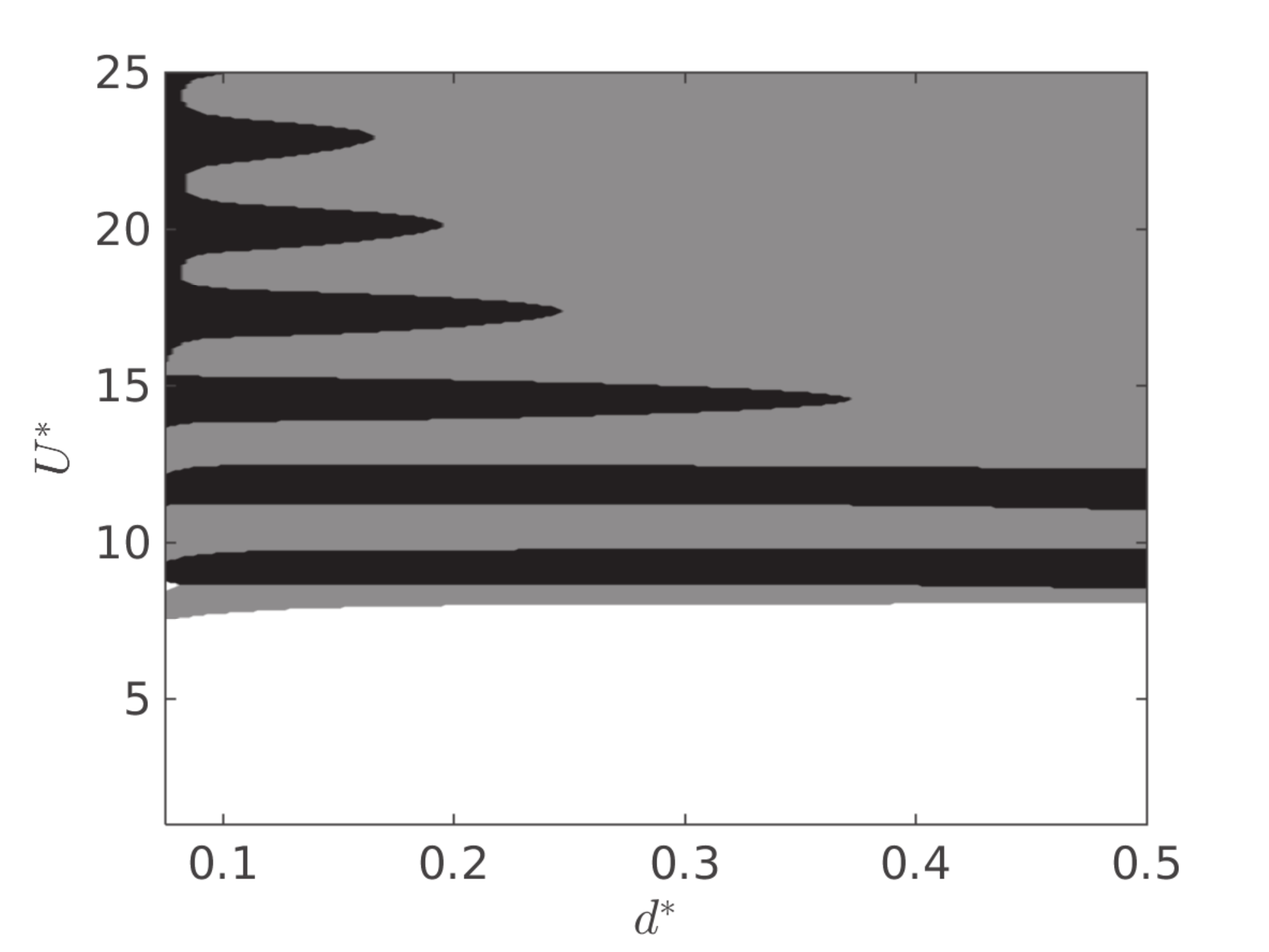} \\
$(a)$ $d^*=0.15$, $M^*=20$ & $(b)$ $H^*=0.1$, $M^*=20$
\end{tabular}
\caption{\label{fig:maps} Effect of $H^*$ $(a)$ and $d^*$ $(b)$ on the synchronization in the most unstable mode. 
Color conventions are identical as in Figure \ref{fig:map_M}.}
\end{figure}

\change{Figure \ref{fig:maps}$(a)$ reveals that out-of-phase modes are predominant when $H^*$ tends to zero while both modes are found for $H^*$ comparable to $d^*$. In addition, comparison with the single-flag threshold shows that the effect of the neighbour becomes small when $H^*\ll 1$, which is consistent with Eq.~\eqref{eq:lin_vji}  and the ${\cal O}({H^*}^2)$ of  the induced velocity scales.
Figure \ref{fig:maps}$(b)$ reveals that in-phase modes are obtained for small values of $d^*$ while out-of-phase modes dominate at larger distances. Noticeably, this latter trend also agrees with two-dimensional experiments and numerical results~\citep{zhang2000,zhu2003}. }

\section{Large-amplitude case}
\label{sec:non_linear}
The numerical \change{approach} presented in more details in Ref.~\cite{michelin2013} is extended here to solve the nonlinear coupled equations for the dynamics of the two flags, Eqs.~\eqref{eq:momentum}--\eqref{eq:inextensibility}, \eqref{eq:potential_ff} and \eqref{eq:lighthill_non_uniform}--\eqref{eq:f_resist}. The beam dynamics are projected along the normal and tangential directions, and the projections on $\boldsymbol{e}_{\tau_i}$ provide the tensions $f_{T_i}$ in both flags which are substituted into the dynamics along the normal directions. 
Using a semi-implicit time-stepping scheme, the entire system then becomes a non-linear equation for $[\partial\theta_1/\partial s_1, \partial\theta_2/\partial s_2]$ solved iteratively at each time step using Broyden's method. Eventually, flapping characteristics can be reconstructed from curvature distributions using clamped-free boundary conditions. 
Initially, one flag is straight and a small uniform curvature is imposed on the second flag. This ensures that initial conditions do not promote in-phase or out-of-phase dynamics.

\begin{table}
\centering
 \begin{tabular}{clccc}
                   &  Configuration     & Frequency  & Amplitude  & Synchronisation \\
                    \hline
$U^*=9$      & \textbf{$\mathbf{L_w=0}$, FF approx. }     & {\bf 3.27 }& {\bf 0.0340} & {\bf out-of-phase} \\
                    & $L_w=1$, FF approx.      & 3.24 & 0.0354 & out-of-phase \\
                    & $L_w=0$                          & 3.26 & 0.0336 & out-of-phase \\
                    & $L_w=1$                         & 3.24 & 0.0349 & out-of-phase \\         
                    & $L_w=2$                         & 3.24 & 0.0348 & out-of-phase \\              
                    & LAEBT single flag            & 3.14 & 0.0271 &     -        \\
                    \hline
$U^*=10$    & \textbf{$\mathbf{L_w=0}$, FF approx. }    & {\bf 3.42} & {\bf 0.0766} & {\bf in-phase} \\
                    & $L_w=1$, FF approx.      & 3.44 & 0.0774 & in-phase \\
                    & $L_w=0$                         & 3.43 & 0.0761 & in-phase \\
                    & $L_w=1$                         & 3.44 & 0.0769  & in-phase \\
                    & $L_w=2$                         & 3.44 & 0.0769  & in-phase \\
                    & LAEBT single flag           & 3.49 & 0.0685 & - \\
                    \hline
$U^*=11.7$  & \textbf{$\mathbf{L_w=0}$, FF approx. }     & {\bf 4.56 }& {\bf 0.1078} & {\bf in-phase} \\
                    & $L_w=1$, FF approx.      & 4.57 & 0.1076 & in-phase \\
                    & $L_w=0$                         & 4.57 & 0.1079 & in-phase \\
                    & $L_w=1$                         & 4.58  & 0.1077 & in-phase \\
                    & $L_w=2$                         & 4.58  & 0.1078 & in-phase \\
                    & LAEBT single flag           & 4.68 & 0.1065 & - \\
                    \hline
  \end{tabular}
\caption{\label{table} Influence of the model hypotheses on the flapping characteristics for $H^*=0.1$, $d^*=0.15$ and $M^*=27$. For both $U^*$, the first line corresponds to cases shown on Figure \ref{fig:LAEBT_dynamics}.}
\end{table}

\begin{figure}
\centering
\includegraphics[width=0.83\textwidth, trim = 0cm 0.78cm 0cm 0cm, clip]{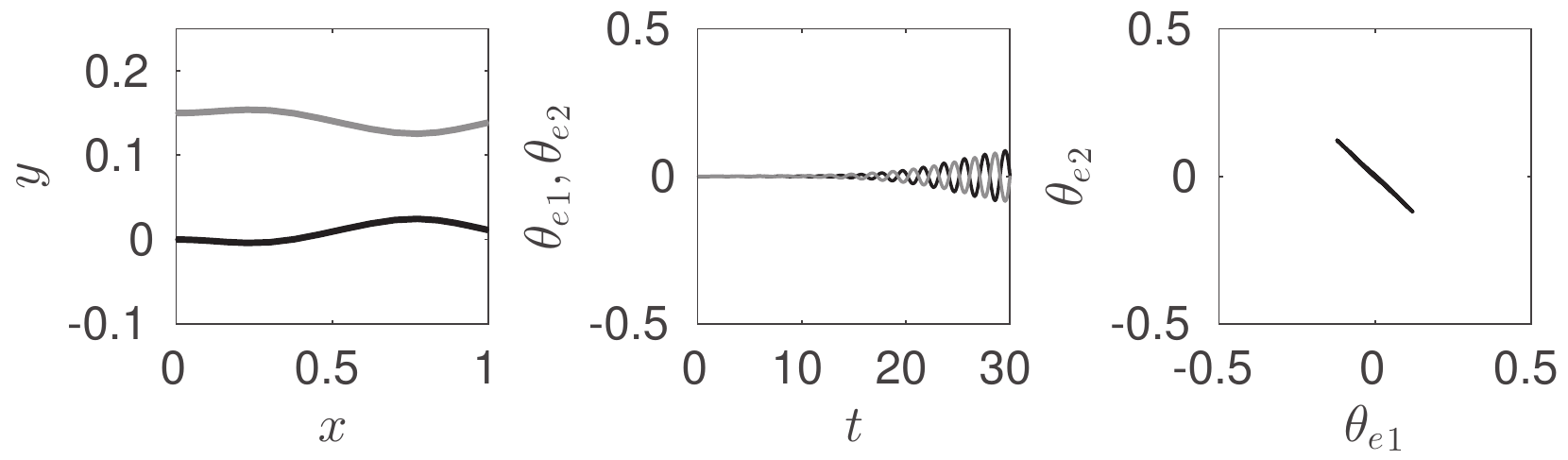}\\
\includegraphics[width=0.83\textwidth, trim = 0cm 0.58cm 0cm 0cm, clip]{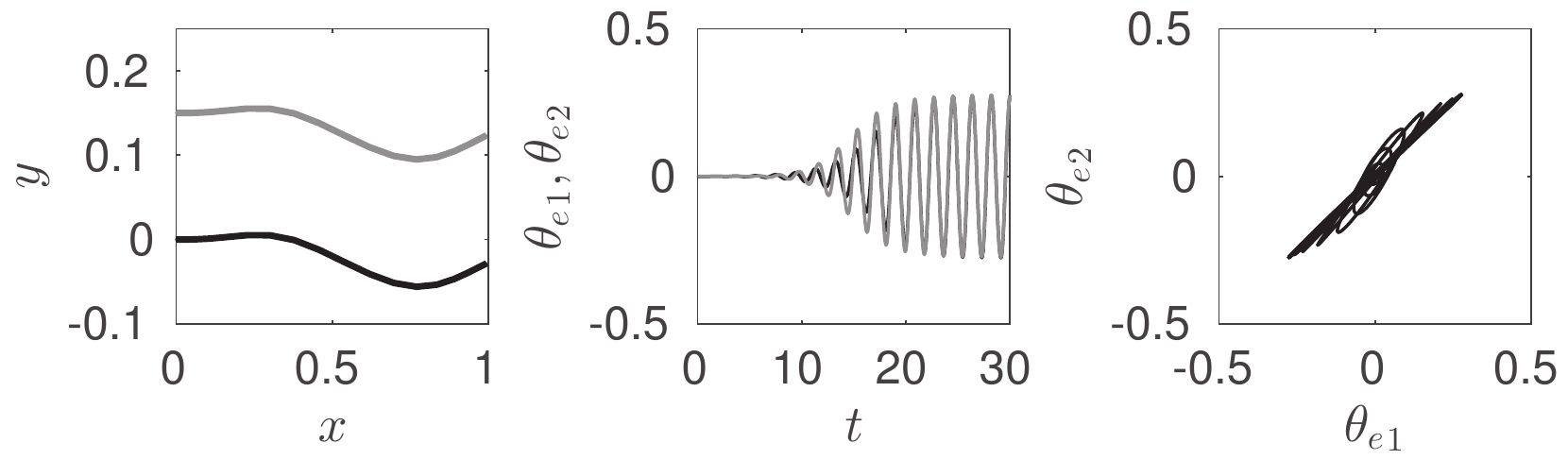}\\
\hspace{-0.3cm}
\includegraphics[width=0.83\textwidth, trim = 0cm 0cm 0cm 0cm, clip]{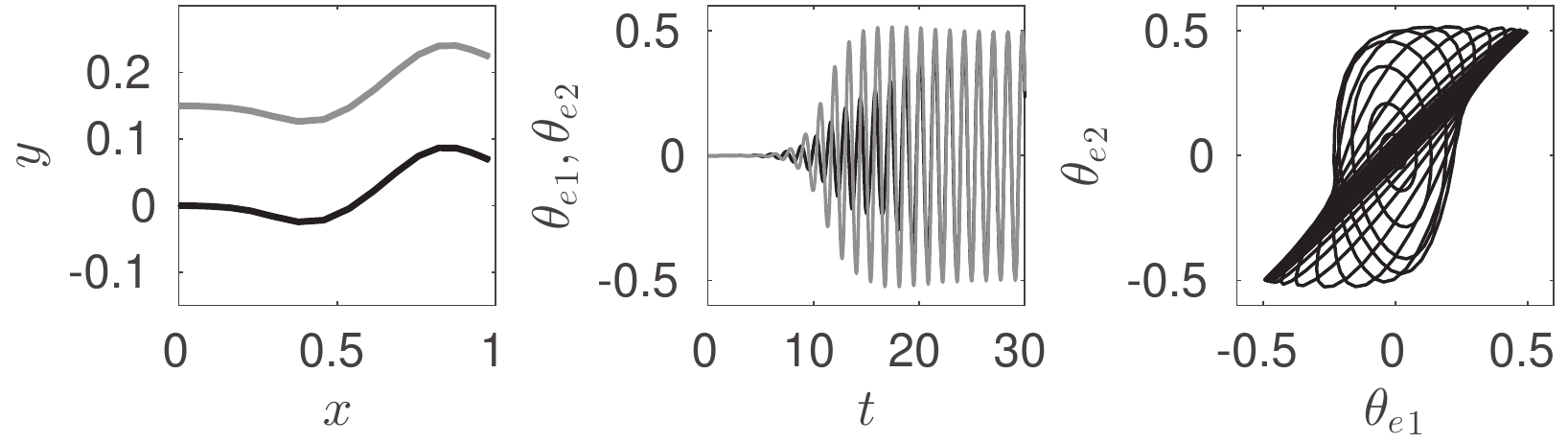} 
\vspace{-0.2cm}
\caption{\label{fig:LAEBT_dynamics} Flapping dynamics for $H^*=0.1$, $d^*=0.15$, $M^*=27$ and $U^*=9$ (top), $U^*=10$ (middle) and $U^*=11.7$ (bottom). The left column shows the flags' instantaneous position. Middle and right columns show the evolution of trailing edge angles $\theta_{ei}$.}
\end{figure}

Figure \ref{fig:LAEBT_dynamics} \change{illustrates} the resulting dynamics: the small perturbations grow on the perturbed flag and set the second flag into motion, leading to exponential growth and saturation of both flapping amplitudes. After a transient regime, flags settle in a permanent regime, either in-phase or out-of-phase, with the same flapping amplitude and frequency. 
\change{For larger $d^*$ (not shown), a similar behaviour is found and the time necessary to reach the saturation of the flapping amplitude is roughly independent of $d^*$. However, the time required to reach synchronization (i.e. the time necessary for the relative phase to converge to its long-term value) increases significantly with $d^*$: flags are not necessarily in-phase or out-of-phase when their flapping amplitude saturates, and a long transient regime may be required before reaching synchronization. Saturation and synchronization therefore occur on different time scales. This is consistent with saturation being mainly driven by the flag's own dynamics and synchronization resulting from hydrodynamic coupling: the latter become weaker when $d^*$ is increased. It should be noted nevertheless that the flapping amplitude $A$ (defined as the half of the peak-to-peak amplitude corresponding to the trailing edge's lateral position in the saturated regime) may be slightly modified during the synchronization process (Table \ref{table}).} 

\change{
The flapping characteristics (i.e. the amplitude, frequency and synchronization type) associated with the dynamics presented on Figure \ref{fig:LAEBT_dynamics} are reported in Table \ref{table} (bold), which validates the different approximations made (i.e. neglecting the wake and the far-field approximation). 
In addition the single flag case is also reported for comparison. 
Results show that both wakes and far-field approximation have little influence on the flapping characteristics for the cases considered here. In particular, the influence of both approximations remain small in front of that of hydrodynamic coupling which is estimated through comparison with the single flag case. This validates the present model for the physical parameters considered in Table \ref{table}. }
\begin{figure}
\centering
\includegraphics[width=0.65\textwidth, trim = 0cm 0.3cm 0cm 1cm, clip]{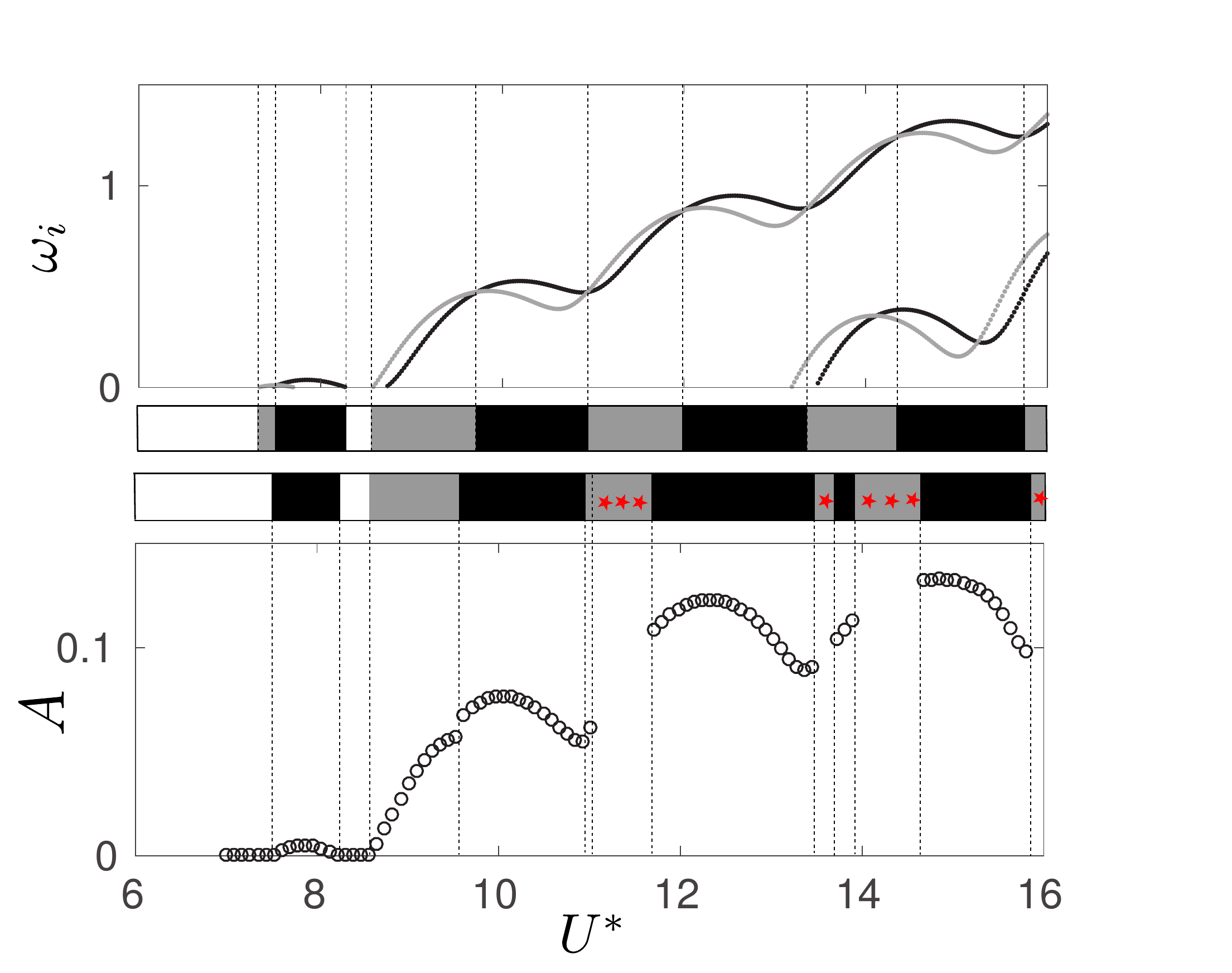}
\caption{\label{fig:LAEBT_M27} Comparison of linear and non-linear results for $H^*=0.1$, $d^*=0.15$ and $M^*=27$. (top) Linear growth rates of in-phase (black) and out-of-phase modes (grey). (bottom) Saturated amplitude obtained from nonlinear simulations. In both cases, the associated colored stripes indicate the flags' synchronization in the most unstable mode and saturated regime: in-phase (black) or out-of-phase (grey). White regions correspond to a stable assembly. Red stars correspond to cases where simulations are stopped due to touching flags.}
\end{figure}

\begin{figure}
\centering
\includegraphics[width=0.55\textwidth, trim = 0cm 0cm 0cm 0cm, clip]{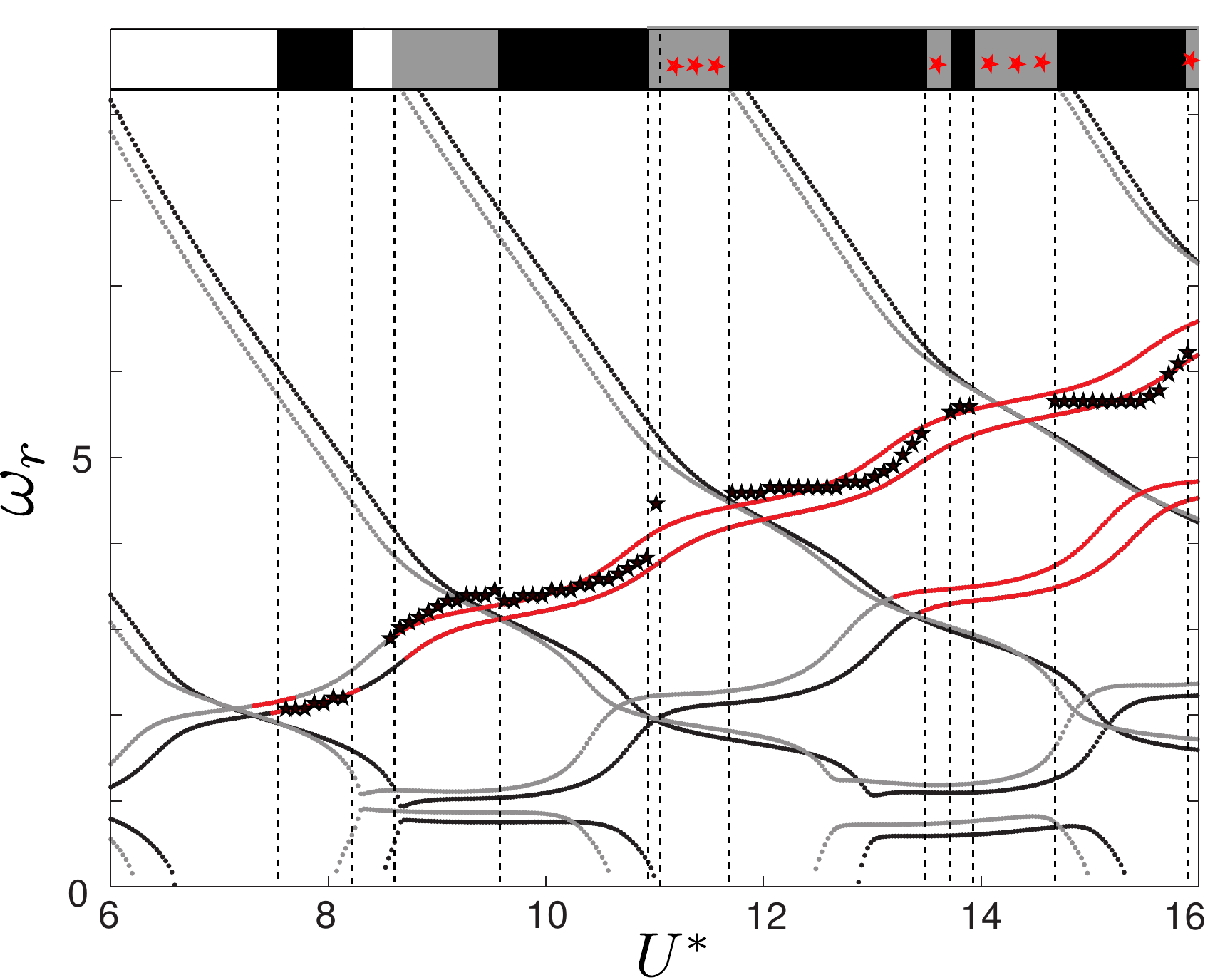}
\caption{\label{fig:LAEBT_M27_freq} Comparison of linear and non-linear frequencies. Linear results correspond to Figure \ref{fig:global_freq}$(b)$, non-linear frequencies in the final state are represented in black stars. Parameters and color conventions are identical as in Figure \ref{fig:LAEBT_M27}. }
\end{figure}

For the same set of parameters, Figures \ref{fig:LAEBT_M27} and  \ref{fig:LAEBT_M27_freq} compare
large amplitude results with the linear predictions.
Figure \ref{fig:LAEBT_M27} displays the linear growth and the saturated amplitude obtained from non-linear simulations. 
As can be seen, final states obtained from LAEBT simulations roughly correspond to the most unstable mode in the linear predictions. The apparent discrepancy found at the lowest threshold ($U^*\approx 7.5$) is due to the very small amplification rate which require long simulation times to get the final saturated state. It has been checked that the out-of-phase area predicted close to threshold in the linear case eventually emerges from LAEBT simulations (but with a very weak amplitude) for very long simulation times. Thresholds are therefore well-reproduced.
Figure \ref{fig:LAEBT_M27_freq} compares the linear frequencies already presented in Figure \ref{fig:global_freq} to the flapping frequencies of the final state obtained from the large amplitude simulations. 
A good agreement is found and large-amplitude results confirm that out-of-phase dynamics is associated with larger frequencies than in-phase dynamics; a trend also reported in the two-dimensional experiments by Ref.~\cite{zhang2000}.  

\begin{figure}
\centering
\includegraphics[width=0.6\textwidth]{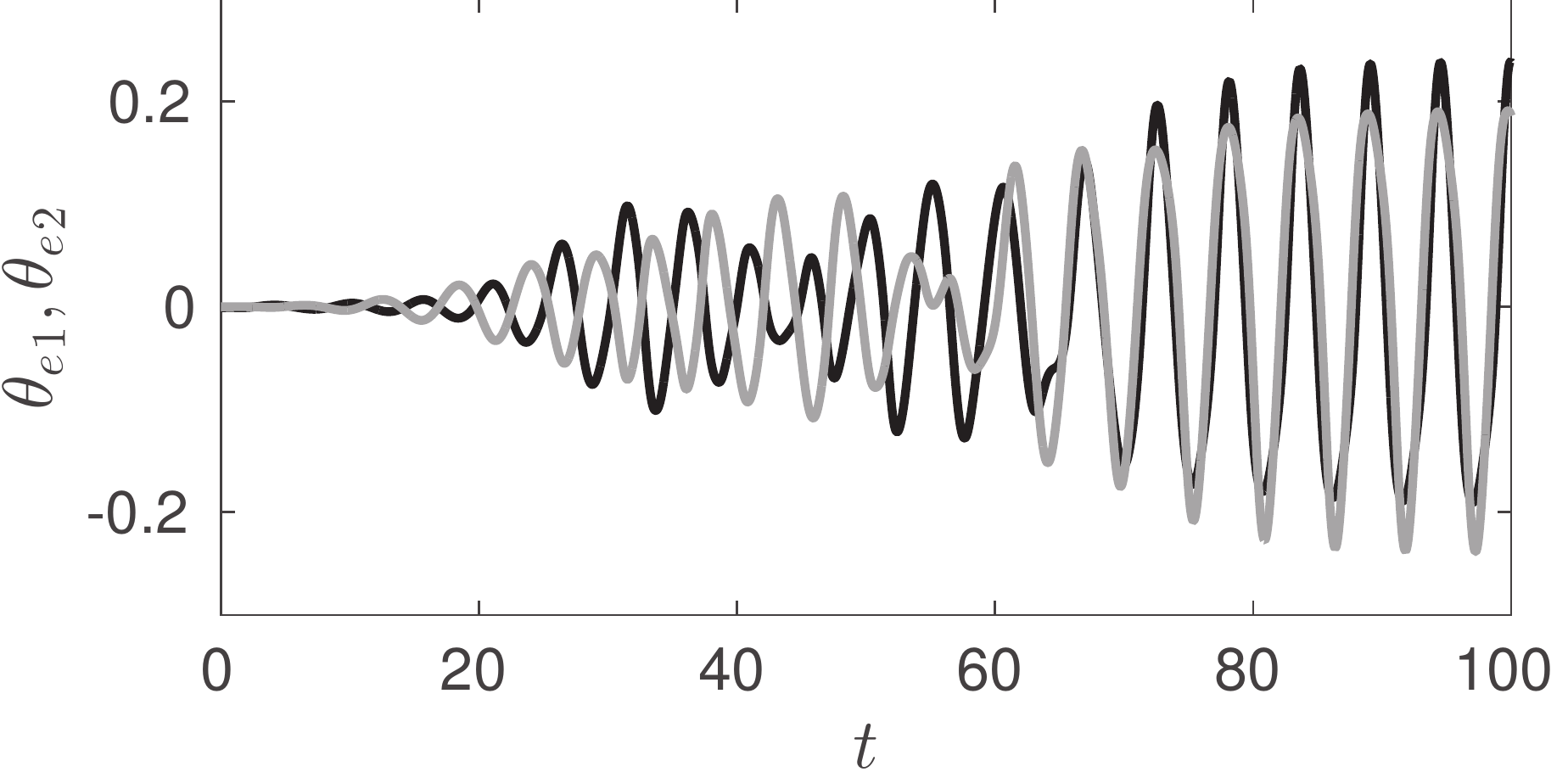}
\caption{\label{fig:LAEBT_dynamics_M3} Trailing edge angles from LAEBT simulation for $M^*=3$, $d^*=0.1$ and $U^*=15$.}
\end{figure}

\change{However, Figure \ref{fig:LAEBT_M27} shows that the non-linear dynamics may differ from linear predictions regarding the flags' synchronization  based on the most unstable eigenmodes  (e.g. for $U^*\approx 11.7$). This behaviour corresponds to Figure \ref{fig:LAEBT_dynamics} (bottom line) where the system is dominated by an out-of-phase mode at the beginning of the transient stage (the out-of-phase mode is most unstable) before non-linearities become important and switch the system to in-phase synchronization. }
This phenomenon is highlighted for an other set of parameters on Figure \ref{fig:LAEBT_dynamics_M3} where the growth of the linear out-of-phase prediction is clearly seen before the system eventually settled in-phase. 
In both cases, the non-linearities therefore appear to favour in-phase synchronization.
This type of non-linear selection is only observed when the most unstable in-phase and out-of-phase modes have similar growth rates, and can be seen as a competition between two modes which is expected to occur frequently for large values of $M^*$ (see Figure \ref{fig:map_M}). 

\section{Conclusion and perspectives}\label{sec:conclusions}
This study proposes a framework to analyze the hydrodynamic coupling of two slender flags in axial flow when the separation distance is small compared to the flag's length but large compared to its width ($H\ll d\ll L$). The essential idea of the present model is to account for hydrodynamic coupling by considering the modifications introduced by each flag in the ambient flow seen by its neighbour. The resulting extensions of Lighthill's Elongated Body Theory (EBT) and Large Amplitude Elongated Body Theory (LAEBT) were used to study the linear and nonlinear dynamics and in particular the role of hydrodynamic coupling in the synchronization of the two flags' dynamics. 
In the linear case, flutter instability leads to either in-phase or out-of-phase modes and hydrodynamic coupling appears to destabilize the system. 
In the nonlinear saturated regime, our simulations show that the flags synchronize after a transient regime. 
The selected flapping dynamics at long times, and in particular the flags' synchronization, generally corresponds to linear predictions and the linear maps are therefore representative of what should be expected in the non-linear regime. Out-of-phase dynamics is thus dominant for small values of $M^*$, while large $M^*$ correspond to a greater sensitivity of the phase to the system's parameters. In addition, in-phase motion is generally expected for small $d^*$ and out-of-phase motion at larger $d^*$. These results are consistent with previous experimental and numerical studies on this topic.

Interestingly, non-linear selection mechanisms are observed to dominate the linear selection in some cases \change{where in-phase and out-of-phase modes have similar growth rates}, and appear to favour in-phase dynamics. 
The mechanisms of this non-linear selection are beyond the scope of the present work and should be the focus of further investigation. Understanding such nonlinear selection mechanisms is an important challenge and still an open question in the domain of fluid-structure interactions. The present problem, and its simplified framework, provide an interesting benchmark configuration to investigate this question in greater depth. 

Finally, the present framework can easily be extended to account for more than two flags~\citep{michelin2009,tian2011b,uddin2013,favier2015} and to couple the fluid-solid system to an electric generator as for piezoelectric flags~\citep{michelin2013,xia2015}.

\begin{acknowledgements}
This work was supported by the French National Research Agency ANR (Grant No. ANR-2012-JS09-0017).
\end{acknowledgements}

\end{document}